\documentclass[twocolumn]{aastex62}

\graphicspath{{./}{figures/}}
\usepackage{gensymb,longtable}
\usepackage{ragged2e}
\accepted{July 28, 2020}
\begin{document}

%--------------  Useful definitions ----------------
\newcommand{\Ha}{\hbox{{\rm H}$\alpha$}}
\newcommand{\Hb}{\hbox{{\rm H}$\beta$}}
\newcommand{\Hg}{\hbox{{\rm H}$\gamma$}}
\newcommand{\HeII}{\hbox{{\rm He}\kern 0.1em{\sc ii}}}
\newcommand{\OII}{\hbox{{\rm [O}\kern 0.1em{\sc ii}{\rm ]}}}
\newcommand{\OIII}{\hbox{{\rm [O}\kern 0.1em{\sc iii}{\rm ]}}}
\newcommand{\CIV}{\hbox{{\rm C}\kern 0.1em{\sc iv}}}
\newcommand{\CIII}{\hbox{{\rm C}\kern 0.1em{\sc iii}{\rm ]}}}
\newcommand{\MgII}{\hbox{{\rm Mg}\kern 0.1em{\sc ii}}}
\newcommand{\FeII}{\hbox{{\rm Fe}\kern 0.1em{\sc ii}}}
\newcommand{\Mbh}{\ensuremath{M_{\rm BH}}} %Mass
\newcommand{\logMbh}{\rm{log}(\Mbh)}
\newcommand{\Lbol}{\ensuremath{L_{\rm bol}}}
\newcommand{\eddratio}{\ensuremath{L_{\rm bol}/L_{\rm Edd}}}
\newcommand{\Mdot}{\ensuremath{\dot{M_{\rm BH}}}} % accretion rate
\newcommand{\lum}{\ensuremath{\lambda L_{3000}}}
\newcommand{\jrt}[1]{\textcolor{red}{#1}}
\newcommand{\yh}[1]{\textcolor{blue}{#1}}
\newcommand{\pyccf}{\texttt{PyCCF}}
\newcommand{\cream}{\texttt{CREAM}}
\newcommand{\jav}{\texttt{JAVELIN}}
\newcommand{\pycecream}{\texttt{PyceCREAM}}
%-----------------------------------------------------

\title{The Sloan Digital Sky Survey Reverberation Mapping Project: \MgII\ Lag Results from Four Years of Monitoring}

\correspondingauthor{Yasaman Homayouni}
\email{yasaman.homayouni@uconn.edu}

\author[0000-0002-0957-7151]{Y. Homayouni}
\affil{University of Connecticut, Department of Physics, 2152 Hillside Road, Unit 3046, Storrs, CT 06269-3046}

\author[0000-0002-1410-0470]{Jonathan R. Trump}
\affiliation{University of Connecticut, Department of Physics, 2152 Hillside Road, Unit 3046, Storrs, CT 06269-3046}

\author[0000-0001-9920-6057]{C. J. Grier}
\affiliation{Steward Observatory, The University of Arizona, 933 North Cherry Avenue, Tucson, AZ 85721, USA}

\author[0000-0003-1728-0304]{Keith Horne}
\affiliation{SUPA Physics and Astronomy, University of St. Andrews, Fife, KY16 9SS, Scotland, UK}

\author[0000-0003-1659-7035]{Yue Shen}
\affiliation{Department of Astronomy, University of Illinois at Urbana-Champaign, Urbana, IL, 61801, USA}
\affiliation{National Center for Supercomputing Applications, University of Illinois at Urbana-Champaign, Urbana, IL, 61801, USA}

\author[0000-0002-0167-2453]{W. N. Brandt}
\affiliation{Dept. of Astronomy and Astrophysics, The Pennsylvania State University, 525 Davey Laboratory, University Park, PA 16802}
\affiliation{Institute for Gravitation and the Cosmos, The Pennsylvania State University, University Park, PA 16802}
\affiliation{Department of Physics, 104 Davey Lab, The Pennsylvania State University, University Park, PA 16802, USA}

\author{Kyle~S.~Dawson}
\affiliation{Department of Physics and Astronomy,
University of Utah, Salt Lake City, UT 84112, USA}

\author{Gloria Fonseca Alvarez}
\affil{University of Connecticut, Department of Physics, 2152 Hillside Road, Unit 3046, Storrs, CT 06269-3046}

\author[0000-0002-8179-9445]{Paul J. Green}
\affiliation{Harvard-Smithsonian Center for Astrophysics, 60 Garden
Street, Cambridge, MA 02138, USA}

\author[0000-0002-1763-5825]{P. B. Hall}
\affiliation{Department of Physics and Astronomy, York University, Toronto, ON M3J 1P3, Canada}

\author[0000-0002-6733-5556]{Juan V. Hern\'{a}ndez Santisteban}
\affiliation{SUPA Physics and Astronomy, University of St. Andrews, Fife, KY16 9SS, Scotland, UK}

\author[0000-0001-6947-5846]{Luis C. Ho}
\affiliation{Kavli Institute for Astronomy and Astrophysics, Peking University, Beijing 100871, China}
\affiliation{Department of Astronomy, School of Physics, Peking University, Beijing 100871, China}

\author[0000-0001-7908-7724]{Karen Kinemuchi}
\affiliation{Apache Point Observatory and New Mexico State
University, P.O. Box 59, Sunspot, NM, 88349-0059, USA}

\author[0000-0001-6017-2961]{C.~S.~Kochanek}
\affiliation{Department of Astronomy, The Ohio State University,
140 West 18th Avenue, Columbus, OH 43210, USA}
\affiliation{Center for Cosmology and AstroParticle Physics, The Ohio
State University, 191 West Woodruff Avenue, Columbus, OH
43210, USA}

\author[0000-0002-0311-2812]{Jennifer I-Hsiu Li}
\affiliation{Department of Astronomy, University of Illinois at Urbana-Champaign, Urbana, IL, 61801, USA}

\author[0000-0001-6481-5397]{B.~M.~Peterson}
\affiliation{Department of Astronomy, The Ohio State University, 140 West 18th Avenue, Columbus, OH 43210, USA}
\affiliation{Center for Cosmology and AstroParticle Physics, The Ohio State University, 191 West Woodruff Avenue, Columbus, OH 43210, USA}
\affiliation{Space Telescope Science Institute, 3700 San Martin Drive, Baltimore, MD 21218, USA}

\author{D. P. Schneider}
\affiliation{Dept. of Astronomy and Astrophysics, The Pennsylvania State University, 525 Davey Laboratory, University Park, PA 16802}
\affiliation{Institute for Gravitation and the Cosmos, The Pennsylvania State University, University Park, PA 16802}

\author{D. A. Starkey}
\affiliation{SUPA Physics and Astronomy, University of St. Andrews, Fife, KY16 9SS, Scotland, UK}

\author[0000-0002-3601-133X]{Dmitry Bizyaev}
\affiliation{Apache Point Observatory and New Mexico State
University, P.O. Box 59, Sunspot, NM, 88349-0059, USA}
\affiliation{Sternberg Astronomical Institute, Moscow State
University, Moscow, Russia}

\author[0000-0002-2835-2556]{Kaike Pan}
\affiliation{Apache Point Observatory and New Mexico State
University, P.O. Box 59, Sunspot, NM, 88349-0059, USA}

\author{Daniel Oravetz}
\affiliation{Apache Point Observatory and New Mexico State
University, P.O. Box 59, Sunspot, NM, 88349-0059, USA}

\author[0000-0002-2364-7240]{Audrey Simmons}
\affiliation{Apache Point Observatory and New Mexico State
University, P.O. Box 59, Sunspot, NM, 88349-0059, USA}

%%%%%%%%%%%%%%%%%%%%%%%%%%%%%%%%%%%%%%
%               Abstract             %
%%%%%%%%%%%%%%%%%%%%%%%%%%%%%%%%%%%%%%

\begin{abstract}
We present reverberation mapping results for the $\MgII\,\lambda2800$\AA\ broad emission line in a sample of 193 quasars at $0.35<z<1.7$ with photometric and spectroscopic monitoring observations from the Sloan Digital Sky Survey Reverberation Mapping project during 2014 - 2017. We find significant time lags between the $\MgII$ and continuum lightcurves for 57 quasars and define a ``gold sample'' of 24 quasars with the most reliable lag measurements. We estimate false-positive rates for each lag that range from 1-24\%, with an average false-positive rate of 11\% for the full sample and 8\% for the gold sample.
There are an additional $\sim$40 quasars with marginal \MgII\ lag detections which may yield reliable lags after additional years of monitoring.
The \MgII\ lags follow a radius -- luminosity relation with a best-fit slope that is consistent with $\alpha=0.5$ but with an intrinsic scatter of 0.36~dex that is significantly larger than found for
the \Hb\ radius -- luminosity relation. For targets with SDSS-RM lag measurements of other emission lines, we find that our \MgII\ lags are similar to the \Hb\ lags and $\sim$2-3 times larger than the \CIV\ lags. This work significantly increases the number of \MgII\ broad-line lags and provides additional reverberation-mapped black hole masses, filling the redshift gap at the peak of supermassive black hole growth between the \Hb\ and \CIV\ emission lines in optical spectroscopy.
\end{abstract}

%%%%%%%%%%%%%%%%%%%%%%%%%%%%%%%%%%%%%%%%%%
%               Introduction             %
%%%%%%%%%%%%%%%%%%%%%%%%%%%%%%%%%%%%%%%%%%
%----------------------------------------
\section{Introduction} \label{sec:intro}
%----------------------------------------
Observations over more than two decades have shown that supermassive black holes (SMBHs) exist at the center of every massive galaxy and that several galaxy properties are correlated with the mass of the central SMBH \citep{Magorrian1998,Gultekin2009,Kormendy2013}. Understanding the ``co-evolution'' of galaxies and their SMBHs, as implied by these correlations, depends critically on accurately measuring SMBH masses over cosmic time.

The masses of nearby SMBHs have been measured using high spatial resolution observations of stellar or gas dynamics \citep[for a review, see][]{Kormendy2013}, or, in one specific case of M87, using the black hole ``shadow'' \citep{EHT_VI}. However, these techniques are not yet possible for higher redshift galaxies (${z\gtrsim 0.3}$) even with next generation facilities. Beyond the local universe, reverberation mapping (RM, e.g. \citealp{Blandford1982, Peterson1993, Peterson2004}) is the primary technique for measuring SMBH masses. Nearly all rapidly accreting SMBHs, observed as quasars or broad-line active galactic nuclei (AGN), exhibit widespread variability on timescales of weeks to years \citep[e.g.][]{MacLeod2012}. RM measures the time lag, $\tau$, between the variability in the continuum and the broad emission lines. 
In the standard ``lamp post" model \citep{Cackett2006}, this time delay is simply the light travel distance between the central SMBH disk and the broad line-emitting region (BLR). 
Assuming that the BLR motion is gravitational
\begin{equation}\label{eq:1}
    \Mbh = \frac{f R_{\rm BLR} \Delta V^2}{G}.
\end{equation}
Determines the virial product where $G$ is the gravitational constant, $R_{\rm BLR} = c\tau$ is the characteristic size of the BLR, $\Delta V$ is the broad emission line width, and $f$ is a dimensionless factor of order unity that depends (in ways still not fully understood) on the orientation, structure, and geometry of the BLR.

Depending on quasar redshift, different emission lines are used to find the correlation between BLR and continuum lightcurves. The Balmer lines \Hb\ and \Ha\ are well-studied in numerous optical RM observations of broad-line AGN at $z<1$ \citep{Peterson1991, Kaspi2000, Peterson2004, Bentz2009, Bentz2010, Denney2010, Grier2012, Barth2015, Du2015, Hu2015,Shen2016a, Du2016a, Du2016b, Grier2017, Pei2017}, with a total of $\sim$100 mass measurements, 
mostly at $z<0.3$.

There are an additional $\sim$60 RM measurements of the \CIV\,$\lambda$1549 emission line for quasars at $z>1.3$ \citep{Kaspi2007,Lira2018, Hoormann2019, Grier2019, Shen2019b}. At intermediate redshifts ($0.7\lesssim z \lesssim 1.5$), $\MgII\,\lambda2800$\AA\ is the strongest broad line in the observed-frame optical.
However, there have been only a handful of successful detections of \MgII\ lags in higher redshift AGN \citep{Shen2016a, Lira2018, Czerny2019}, with many other attempts failing \citep{Trevese2007, Woo2008, Cackett2015}, mostly because the \MgII\ line is generally less variable than the \Hb\ broad line \citep{Sun2015}. The limited number of \MgII\ RM measurements from observed-frame ultraviolet (UV) spectroscopy of nearby AGN show lags that are %loosely constrained but 
broadly consistent with  the \Hb\ lags of the same objects \citep{Clavel1991, Reichert1994, Metzroth2006}.

RM masses over $1 \lesssim z \lesssim 2$ are particularly desirable because these epochs represent the peak of SMBH accretion \citep[e.g. Section 3.2 of][]{Brandt2015}: the current lack of \MgII\ RM measurements fundamentally limits our understanding of %the bulk of 
SMBH growth.

RM studies of local AGN have established a correlation between the \Hb\ broad-line radius and the (host-subtracted) AGN luminosity \citep{Kaspi2000, Bentz2013}. This enables scaling relations to estimate SMBH masses solely from broad-line width and luminosity \citep{Vestergaard2006}. There have been attempts to calibrate \MgII\ single-epoch masses derived from the RM-based \Hb\ radius-luminosity relation in quasars with both broad lines, building analogous single-epoch mass estimators from \MgII\ \citep{McLure2002,Vestergaard2009,Shen2011,Bahk2019}. However, these \MgII\ mass estimators are plagued by bias \citep{Shen2012}, and some aspects of the \MgII\ variability behavior suggest that an intrinsic \MgII\ radius-luminosity relation may not exist \citep{Guo2019}.
Additional RM studies of \MgII\ are critically needed to understand if the \MgII\ line can be used for both single-epoch and RM masses, and in turn if it can be used to complete our understanding of SMBH mass buildup through intermediate redshifts.

In this work we present \MgII\ lag results from four years of spectroscopic and photometric monitoring by the Sloan Digital Sky Survey Reverberation Mapping (SDSS-RM) project. Section 2 describes the details of the SDSS-RM campaign and sample selection criteria, and our methods of time series analysis and lag identification are presented in Section 3. In Section 4 we present tests of lag reliability that motivate our ultimate lag selection criteria and alias removal. Section 5 presents our final lag results, comparing the measured \MgII\ lags with the \Hb\ and \CIV\ lags of the same quasars along with a \MgII\ $R-L$ relation. Finally, we discuss and summarize our work in Section 6. Throughout this work, we adopt a $\Lambda$CDM cosmology with $\Omega_\Lambda = 0.7$, $\Omega_M = 0.3$, and $H_0 = 70$~km~s$^{-1}$~Mpc$^{-1}$.

\begin{figure}[t]
\centering
\includegraphics[width=88mm]{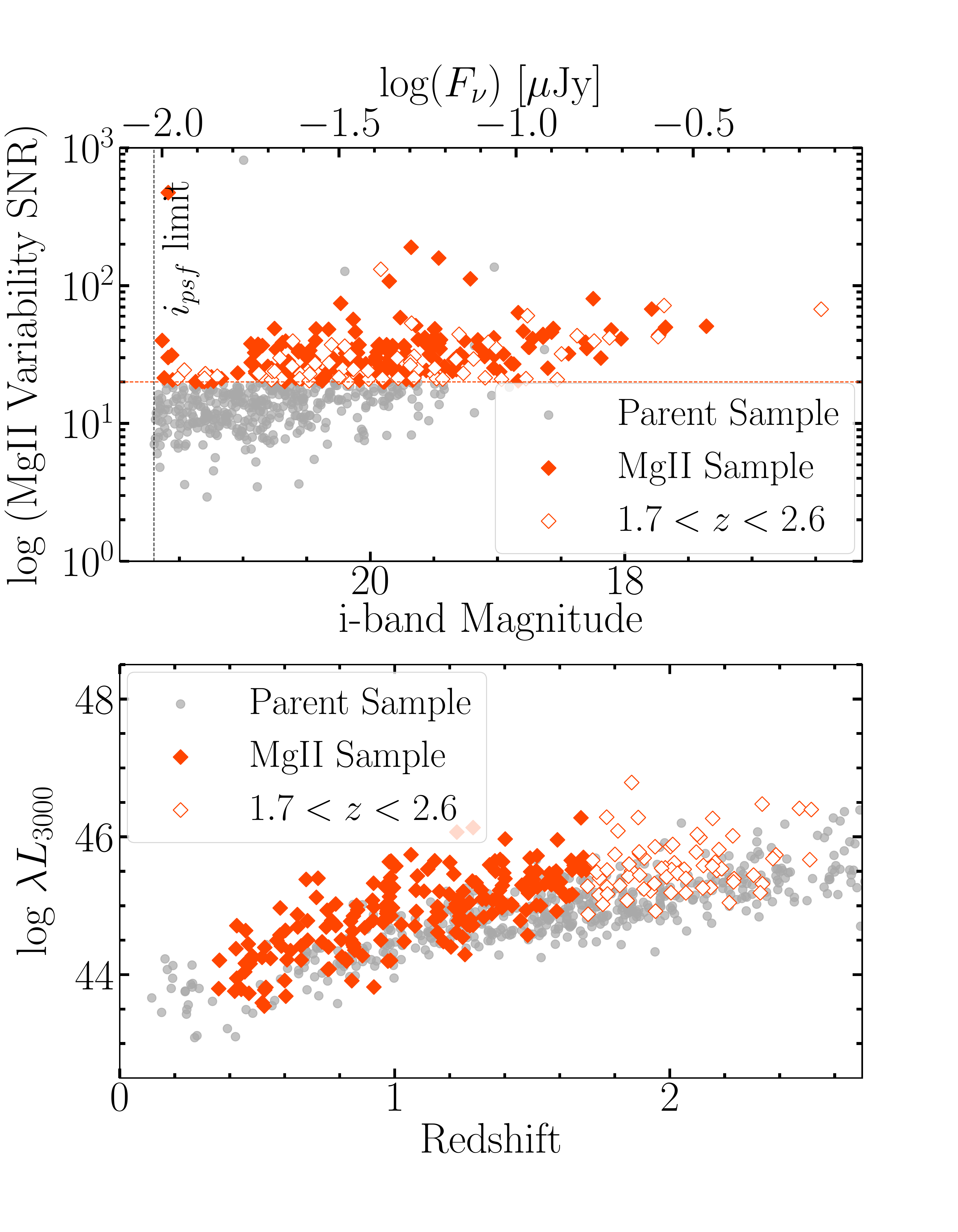}
 \caption{The SDSS-RM parent sample of 849 quasars (gray points) and the \MgII\ subsample of 193 quasars (red filled points). The \MgII\ subsample is selected to have significant \MgII\ variability (top panel; See Section \ref{sec:sample} for more detail) and redshifts within $0.35<z<1.7$ such that \MgII\ is in the observed spectral range and uncontaminated by variable sky emission. The open red symbols show $z>1.7$ quasars where the \MgII\ emission line is variable but frequently affected by telluric contamination. In this paper, all of our analysis is performed on the subsample of 193 targets (filled symbols) with $0.35<z<1.7$.}
\label{fig1}
\end{figure}

\begin{figure}[t]
\centering
\includegraphics[width=80mm]{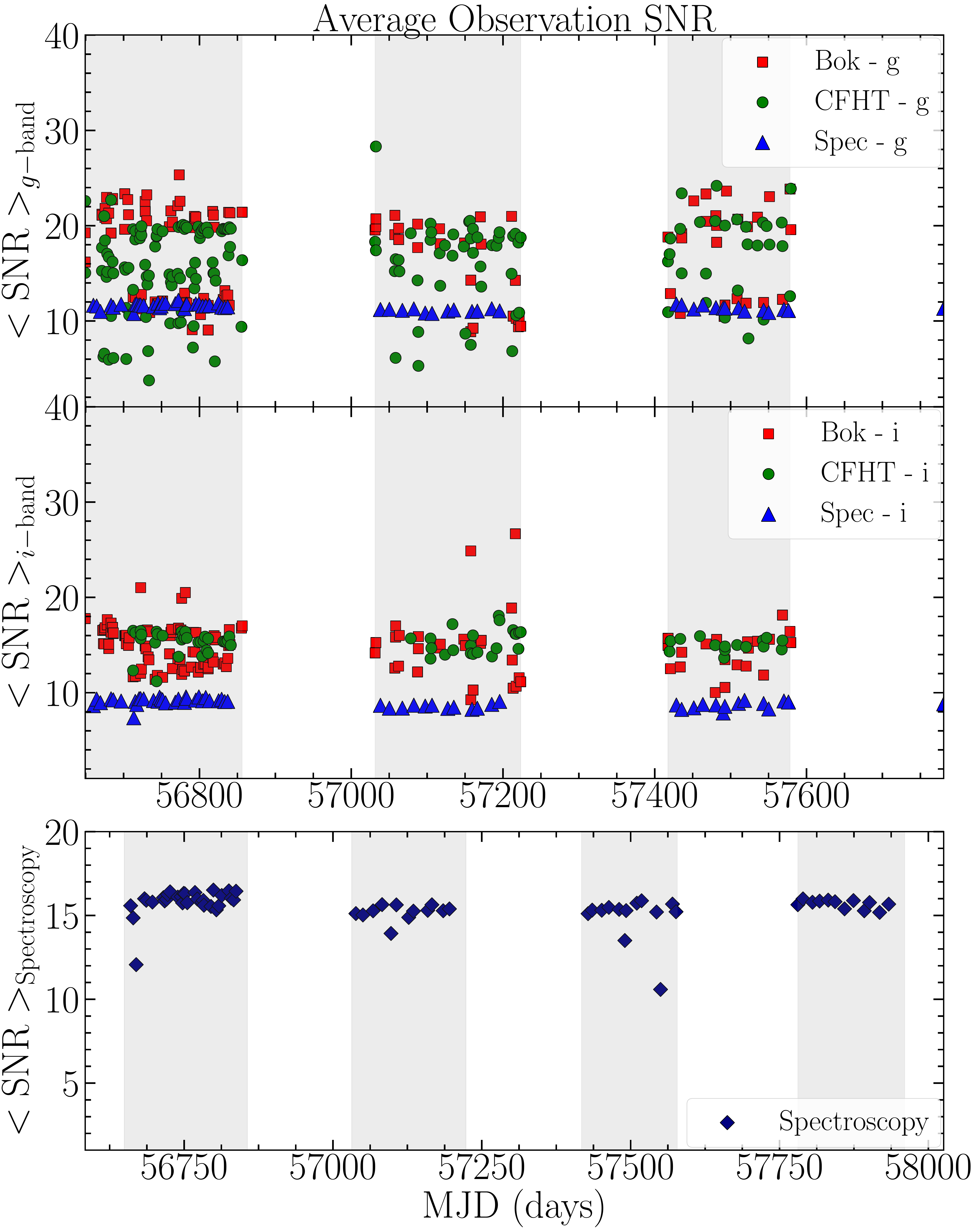}
 \caption{The average SNR and time coverage of the SDSS-RM $g$ (top panel), $i$ (middle panel), and \MgII\ (bottom panel) monitoring observations. SDSS-RM monitors 849 targets every year from Jan to June 2014 (shaded in gray) with a median spectroscopic cadence of 4 days during the first year. Each point represents the average SNR of the \MgII\ emission line for the 193 quasars in our \MgII\ sample observed at that epoch.}
\label{fig2}
\end{figure}

%%%%%%%%%%%%%%%%%%%%%%%%%%%%%
%           Data            %
%%%%%%%%%%%%%%%%%%%%%%%%%%%%%

\section{Data}
%----------------------------------------------
\subsection{Sample Selection}\label{sec:sample}
%----------------------------------------------
Our sample is drawn from the 849 quasars monitored by SDSS-RM, with spectroscopy and photometry in a single 7\,deg$^2$ field observed every year from Jan-Jul since 2014 (see \citealt{Shen2015a, Shen2019a}).
The primary goal of SDSS-RM is to measure lags and black-hole masses for $>$100 quasars spanning a wide range of redshift and AGN properties, using \Hb\ \citep{Shen2016a, Grier2017}, \CIV\ \citep{Grier2019, Shen2019b}, and \MgII\ (\citealp{Shen2016a}; this work). SDSS-RM has also been successful in several related studies of quasar variability \citep{Sun2015, Dexter2019}, quasar emission-line properties \citep{Denney2016a, Shen2016b, Denney2016b, Wang2019}, broad absorption line variability \citep{Grier2016, Hemler2019}, the relationship between SMBH and host galaxy properties \citep{Matsuoka2015, Shen2015b}, and quasar accretion-disk lags \citep{Homayouni2019}. SDSS-RM is a purely magnitude-limited sample ($i_{\rm psf}<21.7$\,mag), in contrast to previous RM studies that selected samples based on quasar variability, lag detectability, and large emission-line equivalent width. This means that SDSS-RM quasars span a broader range of redshift and other quasar properties compared to previous RM studies \citep{Shen2015a}.

To select the targets for this study, we first require that \MgII\ is in the observed-frame optical spectra (i.e., $0.35<z<2.6$). After inspecting the SDSS-RM root-mean-square (RMS) spectra, we found that for $\sim70\%$
%$\sim$30\% for the whole sample (~80 out of 273)
%$\sim70\%$ for the z> 1.7 sample (~49 out of 67 z>1.7)
of the selected targets with $z>1.7$, the \MgII\ line profile is weak with respect to the continuum emission and contaminated by (variable) sky lines, and thus we restrict our parent sample to the 453 quasars with $0.35<z<1.7$.

To ensure that \MgII\ lightcurves are sufficiently variable and have the potential for lag detection, we require a minimum signal-to-noise ratio of the \MgII\ variability, defined as ${\rm SNR2} \equiv \sqrt{\chi^2 - DOF}$. Here $\chi^2$ is the squared deviation of the fluxes relative to the median with respect to the estimated uncertainties,
and ${\rm DOF} = N_{\rm lightcurve} - 1$ is the degrees of freedom of each lightcurve. SNR2 quantifies the deviation from the null hypothesis of no variability, where SNR2 $\sim$\,1 indicates that the variability is dominated by the noise.
%smaller SNR2 values show consistency with the null hypothesis and are less variable while larger SNR2 values indicate highly variable lightcurves.
This quantity is calculated by the PrepSpec \citep{Alard1998} software that is used to flux-calibrate the lightcurves (see Section~\ref{sec:spec} for details). We follow \citet{Grier2019} and require our targets to be significantly variable with ${\rm SNR2}>20$. There are 198 quasars with both $0.35<z<1.7$ and \MgII\ ${\rm SNR2}>20$.
This SNR2 threshold rejects a larger fraction of
\MgII\ targets than it did for the \Hb\ and \CIV\ samples used in \citet{Grier2017, Grier2019}, as \MgII\ is generally less variable than the other strong broad lines in quasars \citep{Sun2015}.

Finally, we reject two targets that have \MgII\ broad absorption lines (BALs)
%['155', '809']
and three targets with weak \MgII\ emission that have average line fluxes consistent with zero.
%['685', '704', '747']
This results in a \MgII\ subsample of 193 quasars in which we search for lags. The properties of these targets are summarized in Figure~\ref{fig1}, and the details of each target are listed in Table \ref{tab:table2}.

%----------------------------------------
\subsection{Spectroscopy}\label{sec:spec}
%----------------------------------------
The SDSS-RM monitoring includes multi-epoch spectroscopy from the BOSS spectrograph \citep{Dawson2013, Smee2013} mounted on the 2.5~m SDSS telescope \citep{Gunn2006}, covering wavelengths of 3650-10400\,\AA\ with a spectral resolution of $R \sim 2000$. We use four years of SDSS-RM spectroscopic observations, obtained annually during dark/grey observing windows from Jan 2014 to Jul 2017 for a total of 68 spectroscopic epochs. During the first year, SDSS-RM obtained a total of 32 epochs with a median cadence of 4 days for the spectroscopy and 2 days for the photometry discussed below, set by weather conditions and scheduling constraints. The following three years had a sparser cadence, with 12 epochs obtained over the 6-month observing window each year. Figure \ref{fig2} shows the median SNR of the continuum and \MgII\ emission line in each epoch for all of the quasars in the \MgII\ subsample. This SNR is computed from the median ratio of the intercalibrated fluxes and the uncertainties (see \ref{sec:lightmerging} for more detail) at each epoch.

The spectroscopic data are initially processed through the standard BOSS reduction pipeline \citep{Dawson2016, Blanton2017}, including flat-fielding, spectral extraction, wavelength calibration, sky subtraction, and flux calibration. The SDSS-RM data are then processed by a secondary custom flux-calibration pipeline that uses position-dependent calibration vectors to improve the spectrophotometric calibrations (see \citealp{Shen2015a} for details). Finally, \texttt{PrepSpec} is used to further improve the relative spectrophotometry and remove any epoch-dependent calibration errors by optimizing model fits to wavelength-dependent and time-dependent continuum and broad-line variability patterns using the fluxes of the narrow emission lines (see \citealp{Shen2016a} for details). \texttt{PrepSpec} also computes a maximum-likelihood SNR for the \MgII\ variability (along with similar variability SNR estimates for the continuum and other emission lines) that is used in our sample selection process (see Section 2.1).

We use the calibrated \texttt{PrepSpec} spectra to compute synthetic photometry in the $g$ and $i$-bands by convolving the calibrated spectra with the SDSS filter response function \citep{Fukugita1996, Doi2010}. The synthetic flux error is computed using the quadratic sum of errors in the measured spectra, errors in the shape of the response function, and the errors in \texttt{PrepSpec} calibration.

To improve the overall quality of the continuum and line light curves, a small number of epochs ($1\%$) are rejected as outliers if offset from the median flux by more than five times the error-normalized median absolute deviation (NMAD). This outlier rejection effectively removes the rare cases of incorrect fiber placement on the SDSS-RM plates.
%--------------------------------------------
\subsection{Photometry}\label{sec:photometry}
%--------------------------------------------
SDSS-RM is supported by ground-based photometry from the 3.6~m Canada-France-Hawaii Telescope (CFHT) MegaCam \citep{Aune2003} and the 2.3~m Steward Observatory Bok telescope 90Prime \citep{Williams2004} imagers. Photometry was obtained in the $g$ and $i$ filters over the full SDSS-RM field, with the same Jan--Jul time coverage over 2014-2017 and a faster cadence than the spectroscopy. The top panels of Figure \ref{fig2} show the average SNR of the $g$ and $i$ flux densities at each photometric epoch for the 193 quasars.

The photometric light curves are extracted from the images using image subtraction as implemented in the ISIS software package \citep{Alard2000}. ISIS aligns all the images and picks a set of images with the best seeing to build a reference image. A scaled reference image, convolved by the point spread function (PSF) at that epoch, is then subtracted from each image to leave only the variable flux. Lightcurves are extracted from the subtracted images and the flux of the quasar in the reference image is added to produce the final lightcurve.

The image subtraction is performed for each individual telescope, filter, CCD and field to produce the $g$ and $i$~lightcurves (Kinemuchi et al. 2020 accepted for publication).

We apply the same outlier rejection method that was implemented on the \MgII\ lightcurves, removing data points that are more than five times the NMAD from the median lightcurve flux. This step excludes data with incorrect photometry due to clouds, nearby bright stars, or detector edges. 

\bigskip

\begin{figure}[t]
\centering
\includegraphics[width=90mm]{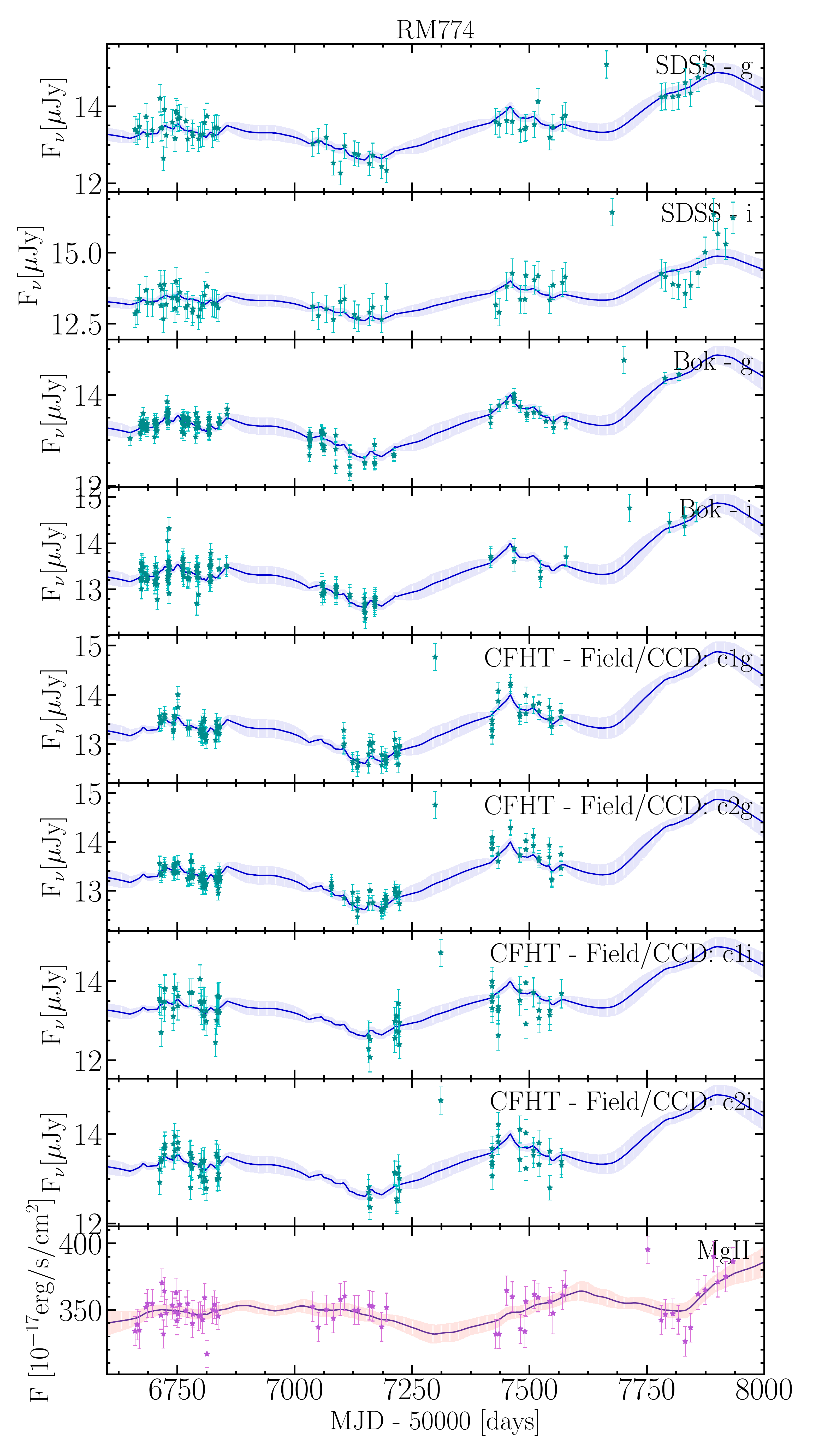}
\caption{A demonstration of lightcurve intercalibation with \texttt{CREAM}, illustrating model fits and rescaled data for the 4-year lightcurves of RMID\,774. Each panel shows the individual pre-merged lightcurve from each observing site in both the $g$ and $i$ filters, the CFHT observations have multiple lightcurves from different fields and CCDs in $\sim$25\% of the sample. The \texttt{CREAM} model prediction and rescaled lightcurves are shown for the post-merged data (cyan for continuum lightcurves and pink for emission-line lightcurve).}
\label{fig3}
\end{figure}
%-------------------------------------------------------
\subsection{Light Curve Merging}\label{sec:lightmerging}
%-------------------------------------------------------
Photometric monitoring using three different observing sites ensures that SDSS-RM has sufficient cadence to produce well-sampled continuum lightcurves. However, combining the multi-site observations requires careful treatment of the differences in seeing, calibration, filter response, telescope throughput, and other site-dependent properties. We use \texttt{CREAM} \citep[Continuum REprocessing AGN Markov chain Monte Carlo;][]{Starkey2016} to inter-calibrate the lightcurves obtained at different sites, following \citet{Grier2017, Grier2019}. 
\texttt{CREAM} models the lightcurves using a power-law prior for the shape of the lightcurve power spectrum, which resembles the observed behavior of AGN lightcurves on short timescales \citep{MacLeod2010, Starkey2016}. To inter-calibrate the lightcurves, the \texttt{CREAM} model is fit to the individual photometric lightcurves from each telescope, filter, and pointing, using a delta-function transfer function and zero lag. Each lightcurve is then rescaled and matched to the model using a multiplicative and additive factor, including rescaled flux uncertainties.

The $g$ and $i$ photometry are merged into a single continuum lightcurve, since the lag between these continuum bands is negligible compared to the expected \MgII\ emission line lags \citep[e.g.,][]{Fausnaugh2016}.
We additionally use \texttt{CREAM} to rescale the \MgII\ lightcurve uncertainties, with extra variance as an additive component and a scale factor as a multiplicative component added in quadrature, while allowing the lag and transfer function to be free parameters. An example of the \texttt{CREAM} lightcurve merging is shown in Figure \ref{fig3}.

Occasionally, the photometric $g$ and $i$ lightcurves are affected by contamination from broad emission-line variability. We computed the 
broad-line variability contamination for the \MgII\ parent sample and identify 4 targets that have $>$10\% contamination in the $g$-band and 5 (different) targets that have $>$10\% contamination in the $i$-band.
%(RM103, RM378, RM600, RM723) in g-band
%(RM371, \textbf{RM400}, RM690, \textbf{RM733}, RM736) in i-band
These broad-line contaminated lightcurves are excluded from the merged continuum lightcurves.

We additionally reject photometric lightcurves from individual pointing/CCDs that are visual outliers compared to the other photometric lightcurves of the same object. These rejected outlier lightcurves are generally associated with imaging problems associated with detector edges, and represent $<$1\% of the observed lightcurves.

%%%%%%%%%%%%%%%%%%%%%%%%%%%%%%%%%%%%%%
%         Time Series Analysis       %
%%%%%%%%%%%%%%%%%%%%%%%%%%%%%%%%%%%%%%
%-----------------------------------------------------------
\section{Time series analysis}\label{sec:lag_identification}
\begin{figure*}[t]
\centering
\includegraphics[width=180mm]{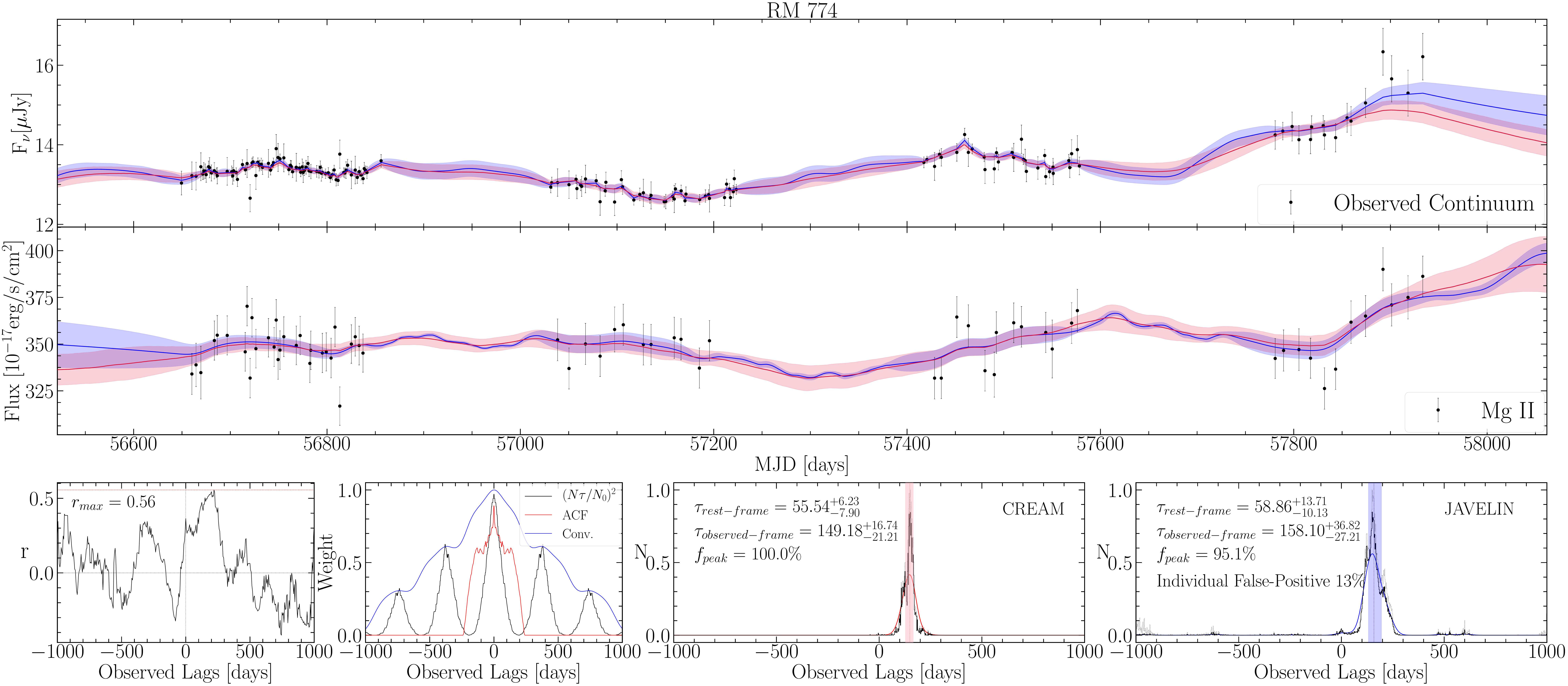}
\caption{Continuum lightcurve (top panel) and \MgII\ lightcurve (middle panel) for RM\,ID\,774 along with lag posteriors (bottom panels). In the top two panels both \texttt{JAVELIN} (blue) and \texttt{CREAM} (red) model fits are shown for the continuum and \MgII\ line lightcurves. The displayed lightcurves are plotted with nightly averages for clarity, although the time-series analysis is computed from the non-averaged observations. \textbf{Bottom Left:} The cross correlation coefficient computed between the continuum and \MgII\ line lightcurve with its maximum displayed by a horizontal red line. \textbf{Second from the left:} The applied weights for our alias removal, with the $[N(\tau)/N(0)]^2$ overlap between lightcurves in black, the continuum auto-correlation function in red, and the final applied weight in blue, obtained from the convolution of the black and red curves. \textbf{Third and fourth from the left:} The unweighted (gray) and weighted (black) lag PDFs computed by \jav\ and \cream. The colored curves indicate the smoothed lag PDFs which are used to find the lag bins. The primary lag is indicated by the colored vertical line with its 16th/84th percentile uncertainties enclosed by the colored shading. Available as a Figure Set for the full sample of 193 AGN.}
\label{fig4}
\end{figure*}

%-----------------------------------------------------------
We measure lags from the SDSS-RM lightcurves following the same approach as \citet{Grier2019}, with two widely used time series analysis methods adapted for multi-year observations: \jav\ \citep{Zu2011} and \cream\ \citep{Starkey2016}.
We do not use the older Interpolated Cross Correlation Function (i.e., ICCF) method \citep{Gaskell1986, Gaskell1987, Peterson2004} that was commonly used in previous RM studies.
ICCF relies on linear interpolation and is less reliable than \jav\ and \cream\ when applied to SDSS-RM and similar RM programs with sparsely sampled monitoring \citep{Grier2017, Li2019}, and ICCF also generally overestimates lag uncertainties \citep{Yu2020}.
For comparison with ICCF lag measurements, we calculate the Pearson coefficient $r$ between the linearly interpolated
continuum and emission line lightcurves (bottom left panel of Figure~\ref{fig4}).

% yh - all the ICCF discussion is commented out for now

%\subsection{ICCF}\label{sec:iccf}
%The Interpolated Cross Correlation Function (i.e., ICCF) has been the most common method applied to previous RM studies \citep{Gaskell1986, Gaskell1987, Peterson2004}. ICCF calculates the Pearson coefficient $r$ between two lightcurves shifted by a range of time lags, using linear interpolation to match the shifted lightcurves onto a uniform grid in time. The most probable time lag is chosen based on the maximum correlation coefficient $r$, and the lag uncertainty is computed from Monte Carlo (MC) iterations of flux resampling and random subset sampling. We implement ICCF using the publicly available \pyccf\ code \citep{Sun2018b} with an interpolation step of 1 day and a lag search range of $\pm$ 500 days. We use 5000 Monte Carlo iterations of flux resampling and random subset sampling for \pyccf\ to generate a cross-correlation centroid distribution (CCCD), which gives the distribution of measured lags in all of MC realizations. The primary peak, with the largest integrated area between local minima, is selected from the CCCD smoothed by 12 days. The final lag is computed from the median of the primary peak using the unbinned CCCD, with upper and lower lag uncertainties measured from the 16th and 84th percentiles of the primary peak.
%----------------------------------
\subsection{JAVELIN}\label{sec:jav}
%----------------------------------
\jav\ \citep{Zu2011} assumes that the quasar variability lightcurve can be modeled by a damped random walk (DRW) process. The DRW description of quasar stochastic variability is well-motivated by observations \citep{Kelly2009, MacLeod2010, MacLeod2012, Kozlowski2016} for the variability timescales probed by SDSS-RM. \jav\ uses a Markov chain Monte Carlo approach using a maximum likelihood method to fit a DRW model to the continuum and emission-line lightcurves, assuming that the line lightcurve is a shifted, scaled, and smoothed version of the continuum lightcurve.

We allow the DRW amplitude to be a free parameter but fix the DRW damping timescale to 300 days since this quantity is not well constrained by the SDSS-RM monitoring duration. We also tested damping timescales of 100, 200 and 500 days and found no significant differences in the measured lags (as expected; e.g. \citealt{Yu2020}). The response of the line lightcurve is parameterized as a top-hat transfer function, assuming a lag and scale factor that is a free parameter with a fixed transfer function width of 20~days. Our observations are not sufficient to constrain the width of transfer-function, resulting in unphysical transfer function widths if left as a free parameter in \jav. A 20-day transfer function width is sufficiently short compared to the expected lag. We tested transfer function widths of 10 and 20~days, motivated by velocity resolved lag observations \citep{Grier2013, Pancoast2018}, with no significant differences in the measured lags. A broader transfer function width of 40 days resulted in significantly different lags for only $\sim$10\% of our sample. We adopt a lag search range of $\pm$1000 days, chosen to be less than the $\sim$1300~day monitoring duration from Jan~2014 to Jul~2017. \jav\ returns a lag posterior distribution from 62500 MCMC simulations which is used to compute the lag and its uncertainty.

%----------------------------------
\subsection{CREAM}\label{sec:cream}
%----------------------------------
\cream\ \citep{Starkey2016} models the driving lightcurve variability with a random walk power spectrum prior $P(f)\propto f^{-2}$, motivated by the lamp post model \citep{Cackett2007}.
The observed continuum lightcurves are only a proxy for the ionizing continuum, and so \cream\ constructs a new driving lightcurve and models both the observed continuum and line emission as smoothed versions of this ionizing continuum model.
\cream\ fits a top-hat response function to the emission-line lightcurve, returning a lag posterior probability distribution while simultaneously inter-calibrating the lightcurves. 

Here we use a \texttt{Python} implementation of \cream\ called \pycecream\footnote{https://github.com/dstarkey23/pycecream}.
We adopt a high frequency variability limit of 0.3 cycles per day and normal priors of $\mathcal{N}$(1.2, 0.2) for the multiplicative error rescaling parameter and normal priors of $\mathcal{N}$(0.5, 0.1) for the variance expansion parameter. As with \jav, we allow \cream\ to probe a lag search range of $\pm$1000~days.

%%%%%%%%%%%%%%%%%%%%%%%%%%%%%%%%%%%%%%%%%%
%          lag selection criteria         %
%%%%%%%%%%%%%%%%%%%%%%%%%%%%%%%%%%%%%%%%%%
\section{Lag Reliability \& Significance}
%-----------------------------------------------
\subsection{Lag Identification \& Alias Removal}
%------------------------------------------------
The posterior lag distributions from \jav\ or \cream\ occasionally contain a primary peak accompanied by other less-significant peaks. The presence of multiple peaks in the posterior lag distribution, also known as aliasing, is a potential outcome of lag detection with sparse sampling data. Aliasing can be caused by matches of weak variability features between the continuum and line lightcurves, because the lag detection MCMC algorithm does not converge, and/or by quasi-periodic variations. The presence of seasonal gaps in multi-year RM data might also cause the lag detection algorithm to inappropriately prefer lags that fall in seasonal gaps where the lightcurve is interpolated with the DRW model prediction in \jav\ or \cream\ rather than directly constrained by observations.

To address the aliasing, we adopt the same lag identification and alias removal procedures of \citet{Grier2019} based on applying a weight to the posterior lag distribution. The weight prior avoids aliased solutions by penalizing parts of the lag posterior that have little overlap between the observed continuum and emission-line lightcurves. This ensures that the final lag search range and lag uncertainties correspond to observationally-motivated lags.

There are two components to the weight prior. For the first component we use the number of overlapping observed epochs between each target's continuum and line lightcurve, given a time lag $\tau$.
If this lightcurve shift results in fewer overlaps between the observed continuum and line lightcurves (e.g., time lags of $\sim$180~days), it is less probable for the lag to be recovered, while more overlapping data points lead to a more secure lag detection. Following \citet{Grier2019} we adopt the overlapping probability weight $P(\tau) = [N(\tau)/N(0)]^2$, where $N(\tau)$ corresponds to the number of overlapping continuum lightcurve and $\tau$-shifted line lightcurve points and $N(0)$ is the number of overlapping data points with no lag, i.e., $\tau = 0$. We force the weight prior to be symmetric by computing $P(\tau)$ for the line lightcurve shifted by $\tau>0$ with respect to the continuum and then assigning the same values at $\tau=-\tau$.

The second component of the weight prior uses the auto-correlation function (ACF) as a measure of how the continuum variability behavior affects our ability to detect lags. For example, a narrow auto-correlation function indicates rapid variability, in which case seasonal gaps are likely to have consequential effects on our lag detection sensitivity. The final weight prior is the convolution between the overlapping probability, $P(\tau)$, and the continuum lightcurve ACF (forcing $\mathrm{ACF} = 0$ when it drops below zero). We refer to the application of the final weight to the posterior lag distributions of \jav\ and \cream\ as the \textit{weighted} lag posteriors.

To identify the time lag from the weighted posterior lag distribution we first smooth the weighted posteriors by a Gaussian filter with a width of 12 days, which helps to identify the peaks in the weighted lag posteriors. The primary peak in the weighted and smoothed lag posteriors are identified from the peak with the largest area, and smaller ancillary peaks in the lag posterior are considered insignificant for our lag identification. Within this primary peak, the expected lag, $\tau$, is determined from the median of the \textit{unweighted} lag posteriors and the lag uncertainty is calculated from the 16$^{th}$ and 84$^{th}$ percentiles. Figure \ref{fig4} provides an example of our alias removal approach and lag detection. 

\begin{figure}
\centering
\includegraphics[width=88mm]{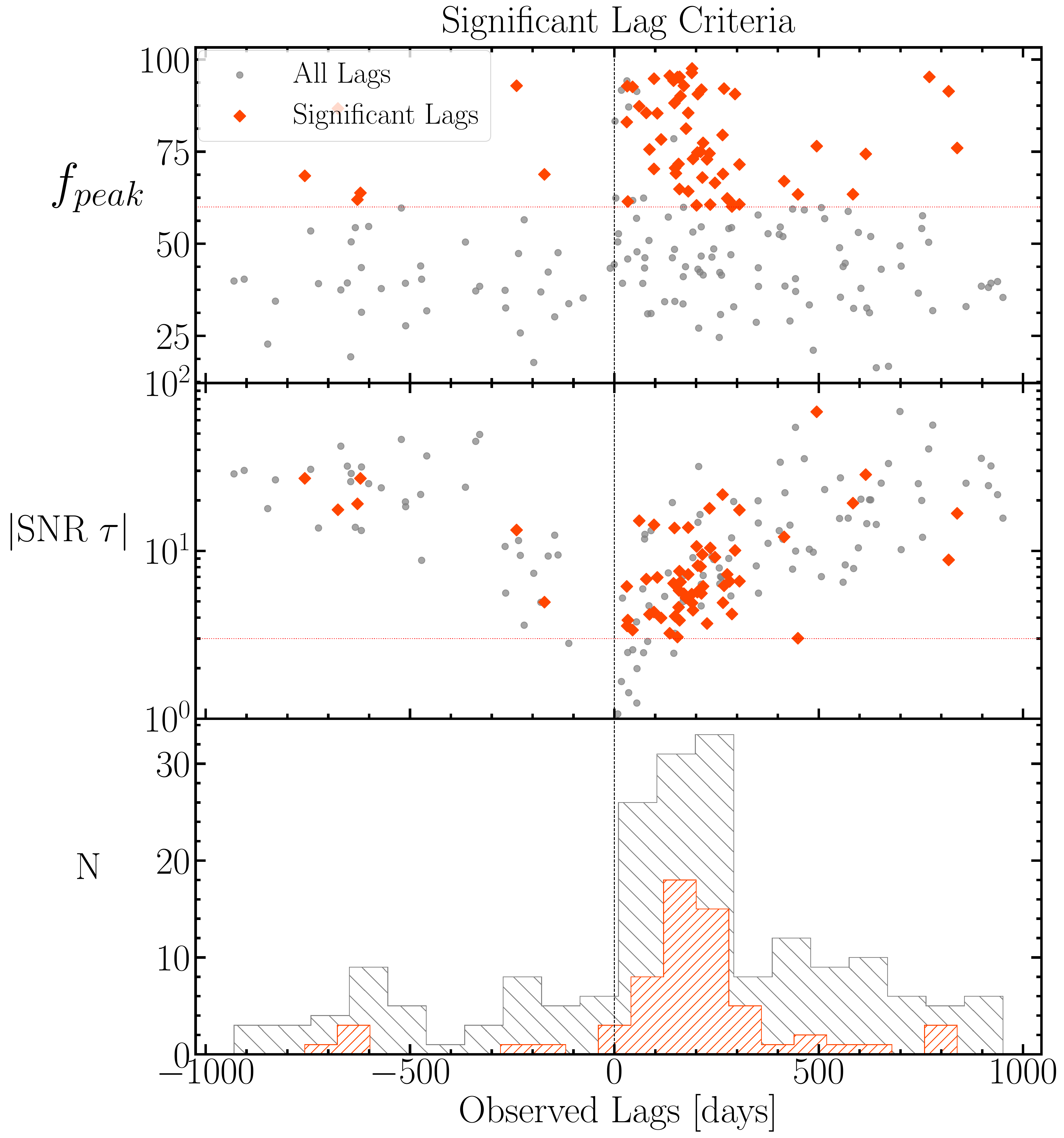}
\caption{The lag significance criteria for the \jav-measured lags.
\textbf{Top:} The fraction of the lag posterior within the primary peak, $f_{peak}$.
\textbf{Middle:} The absolute value of the lag SNR.
\textbf{Bottom:} The histogram of measured lags for the full sample of 193 AGN (gray) and the sample of significant lags (red). The sample includes 57 significant and positive lags that meet both the $f_{peak}$ and $|\mathrm{SNR}|$ criteria (red lines in the top 2 panels, defined in Section 4.2), with an average false-positive rate of 11\%.}
\label{fig5}
\end{figure}
%-----------------------------------------------------------
\subsection{``Significant" Lag Criteria}\label{sec:lag_sig}
%-----------------------------------------------------------
Our lag identification approach removes many secondary peaks and aliases. We require several additional criteria to ensure the final reported lags are statistically meaningful, following
a similar approach to \citet{Grier2019}.
%% ICCF related discssions:
%% and add criteria to select statistically-motivated lags. 
%%One of the criteria we consider is the Pearson correlation coefficient, $r$, that requires correlation between continuum and line lightcurve behaviors. Upon examining our lightcurves we adopt a threshold of 0.4; $r_{max}>0.4$ within $\pm 1 \sigma$ of the measured lag. 
The first criterion is to require that 60\% of the weighted lag posteriors samples are within the primary peak, i.e. $f_{\rm peak} > 60\%$. The primary peak, defined in the previous subsection, is the region of the smoothed lag posterior between local minima with the largest area. The $f_{\rm peak}$ requirement ensures a reliable lag solution and removes cases with many alias lags in the posterior.
We also require significant lags to be well-detected as 3$\sigma$ different from zero, $|\tau|>3\sigma_\tau$.

In summary, our criteria for statistically meaningful lags are:
\begin{itemize}
    \item $f_{peak} > 60\%$: A primary lag peak that includes at least 60\% of the weighted lag posterior samples.
    \item $|\rm SNR(\tau)| > 3$: Minimum of 3$\sigma$ difference from zero lag between the absolute value of the measured lag and its uncertainty. If the lag is positive the noise is the lower-bound uncertainty and if the lag is negative the noise is the upper-bound uncertainty.
\end{itemize}

Figure \ref{fig5} shows the lag-measurement results for all 193 of our targets. The lag-significance criteria are shown in each panel. There are 63 \MgII\ lags that meet the significant lag criteria, with 57 positive lags (shown as red points in Figure~\ref{fig5}). Table~\ref{tab:table2} reports the properties of these 57 quasars, drawn from \citet{Shen2019a}.

\begin{figure}[t]
\centering
\includegraphics[width=85mm]{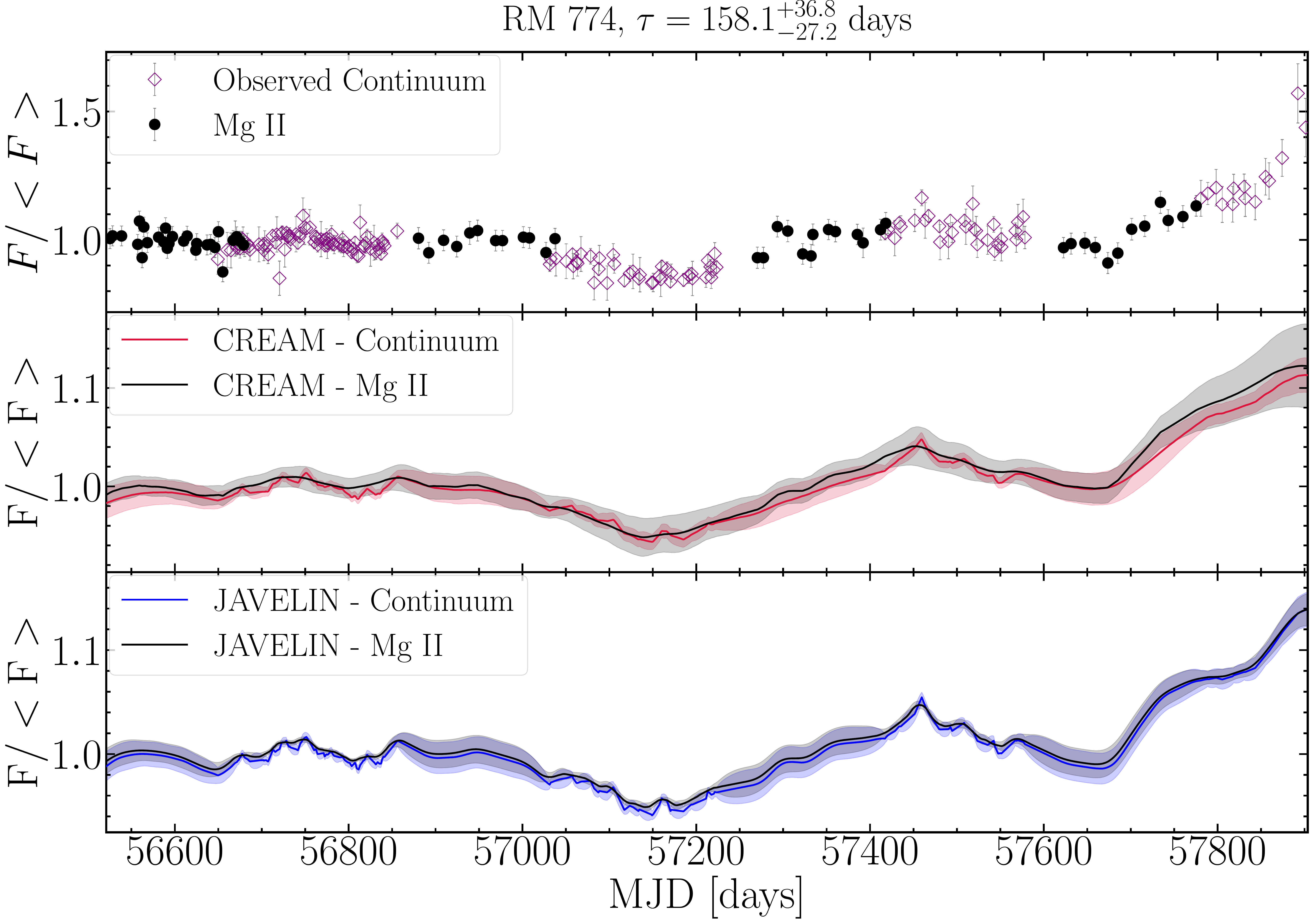}
 \caption{The overlapping continuum and \MgII\ lightcurves and best-fit \cream\ and \jav\ models for RM~774, with the \MgII\ lightcurve shifted by the measured lag. In this example the lag is 158~days, such that the shifted \MgII\ observations fall within the seasonal gap of the continuum observations. However, the lag remains significant and well-constrained because the lightcurve has slow variations on multi-year timescales, such that the lag corresponds to periods in which both the continuum and shifted-\MgII\ lightcurves are both varying in low or high flux states. Available as a Figure Set for the sample of 57 significant positive lags.}
\label{fig:overlaplc}
\end{figure}

As an additional check on the measured lags, Figure~\ref{fig:overlaplc} presents the overlapping continuum and lag-shifted \MgII\ lightcurves and the \cream\ and \jav\ model fits. The overlapping lightcurves are especially instructive for lags of $\sim$180~days in which the shifted \MgII\ observations fall in the seasonal gap of the photometric observations, casting doubt on the reliability of the lag detection. In general these lags are associated with lightcurves that have smooth, low-frequency variations on multi-year timescales, like the example shown. In such cases the lag posterior is well-constrained with a strong primary peak corresponding to when both the continuum and shifted \MgII\ lightcurves are in low or high flux states. Significant lag detections of $\sim$180~days can only be found for slow-varying lightcurves like the example shown in Figure \ref{fig:overlaplc}. Lightcurves with variations on short timescales (i.e. high-frequency variability) require more overlap between shifted lightcurves for significant lag detection. Similar results have also been reported by \citet{Shen2019b} for \CIV\ lightcurves.
\bigskip

%--------------------------------------------------------------------------
\subsection{Rate of False-Positive Lags and ``Gold Sample"}\label{sec:fpr}
%--------------------------------------------------------------------------

Large RM survey programs like SDSS-RM will inevitably include some number of false-positive lag detections. In particular, the limited cadence and seasonal gaps might allow for %\jav\ \& \cream\
lag PDFs with well-defined peaks that meet our significant lag criteria but result from superpositions of non-reverberating lightcurves rather than genuine reverberation. We estimate the average false-positive rate of our lag detections by using the fact that our lag detection analysis does not include any preference for positive versus negative lags, with a lag search range and weighted prior that are both symmetric over $-1000<\tau<1000$~days. If the sample included only non-reverberating lightcurves and lag detections from spurious overlapping lightcurves, the number of positive and negative lag detections would be equal. On the other hand, genuine broad-line reverberation should produce only positive lags.

Our sample includes a total of 6 negative and 57 positive lags that meet the significance criteria defined in Section 4.2. The negative lags are likely the result of spurious lightcurve correlations rather than broad-line reverberation, and the symmetric nature of our lag analysis means there is likely a similar number of spurious positive lags. Thus we use the ratio of negative to positive lag detections as an estimate of the average false-positive rate: with 6 negative and 57 positive lags, the false-positive rate is 11\%.

Figure \ref{fig5} demonstrates that our sample includes significantly more positive than negative lags even for lags below our significance criteria ($f_{\rm peak}>60\%$ and $|\mathrm{SNR}(\tau)|>3$). In the full sample, there are 149 positive and 44 negative lags, indicating an overall false-positive rate of 30\%. %\yh{Out of the 105 true positive lags, we identify 57 significant, positive lags and 48 lags that are less significant}.
The larger number of positive lags in the full sample indicates that an additional 40-50 of the positive lags are likely to be true positive lags. Many of these lower-significance positive lags are likely to become significant detections with additional SDSS-RM monitoring planned as part of the SDSS-V survey \citep{Kollmeier2019}.

\begin{figure}[t]
\centering
\includegraphics[width=85mm]{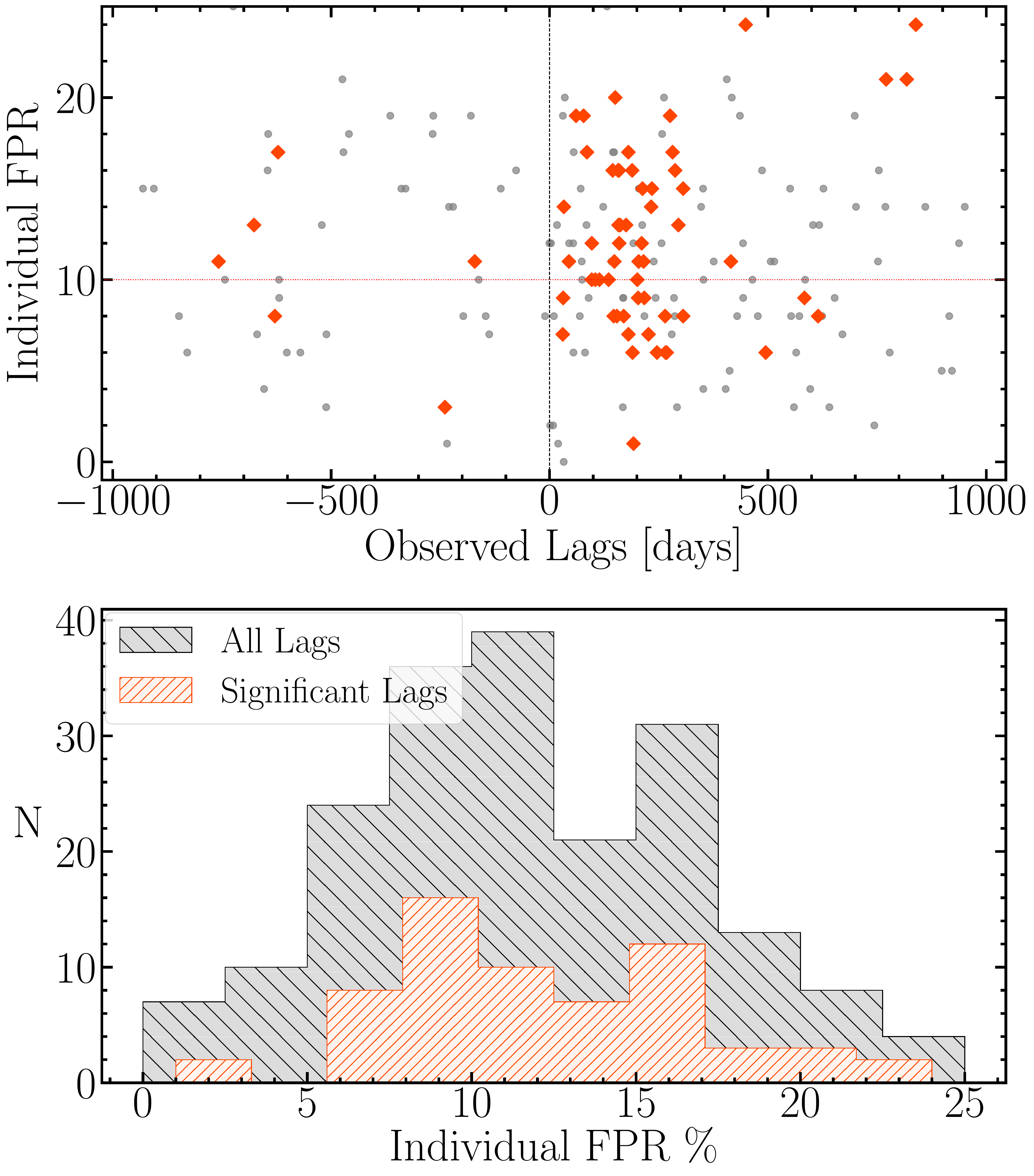}
\caption{\textbf{Top:} The distribution of individual false-positive rates and measured lags for the full sample (gray) and significant lags (red). False positive rates are measured from matching each quasar continuum lightcurve with \MgII\ lightcurves of different quasar, repeated 100 times.
\textbf{Bottom:} A histogram of false-positive rates measured for full sample (gray) and the sample of significant lags (red). The significant lag sample has an average false-positive rate of 12\% from this method, with a ``gold sample'' of 24 significant and positive lags with false-positive rates of $\le$10\%.}
\label{fig:lagfpr}
\end{figure}

The false-positive rate measured from the ratio of negative to positive lags is a robust indication of the overall sample reliability. However, not all lags in our sample are equally likely to correspond to physical reverberation or spurious correlations. To address this, we design an individual false-positive rate test on all 193 set of lightcurves as a measure of each lag's likelihood of being true. We measure \jav\ lag posteriors from each AGN continuum lightcurve matched to the \MgII\ lightcurve of a different AGN, repeating this process 100 times (and excluding duplications). Since the lightcurves from different AGN are uncorrelated, any lag detections meeting our significance criteria are false positives. The individual false-positive rates for the 57 positive significant lags are reported in Table~\ref{tab:table2} and shown in Figure \ref{fig:lagfpr}. The average of the individual false-positive rates for the 57 positive significant lags is 11\%, similar to the 11\% false-positive rate for the sample measured from the ratio of significant negative to positive lags.

We use the individual false-positive rates to define a ``gold sample'' of the most reliable lag measurements with individual false-positive rates of $\le$10\%. The gold sample includes 24 significant, positive \MgII\ lags.
%--------------------------------------------------------------------------
\subsection{Lag Comparison: JAVELIN and CREAM}\label{sec:method_comparison}
%--------------------------------------------------------------------------

\begin{figure}[t]
\centering
\includegraphics[width=80mm]{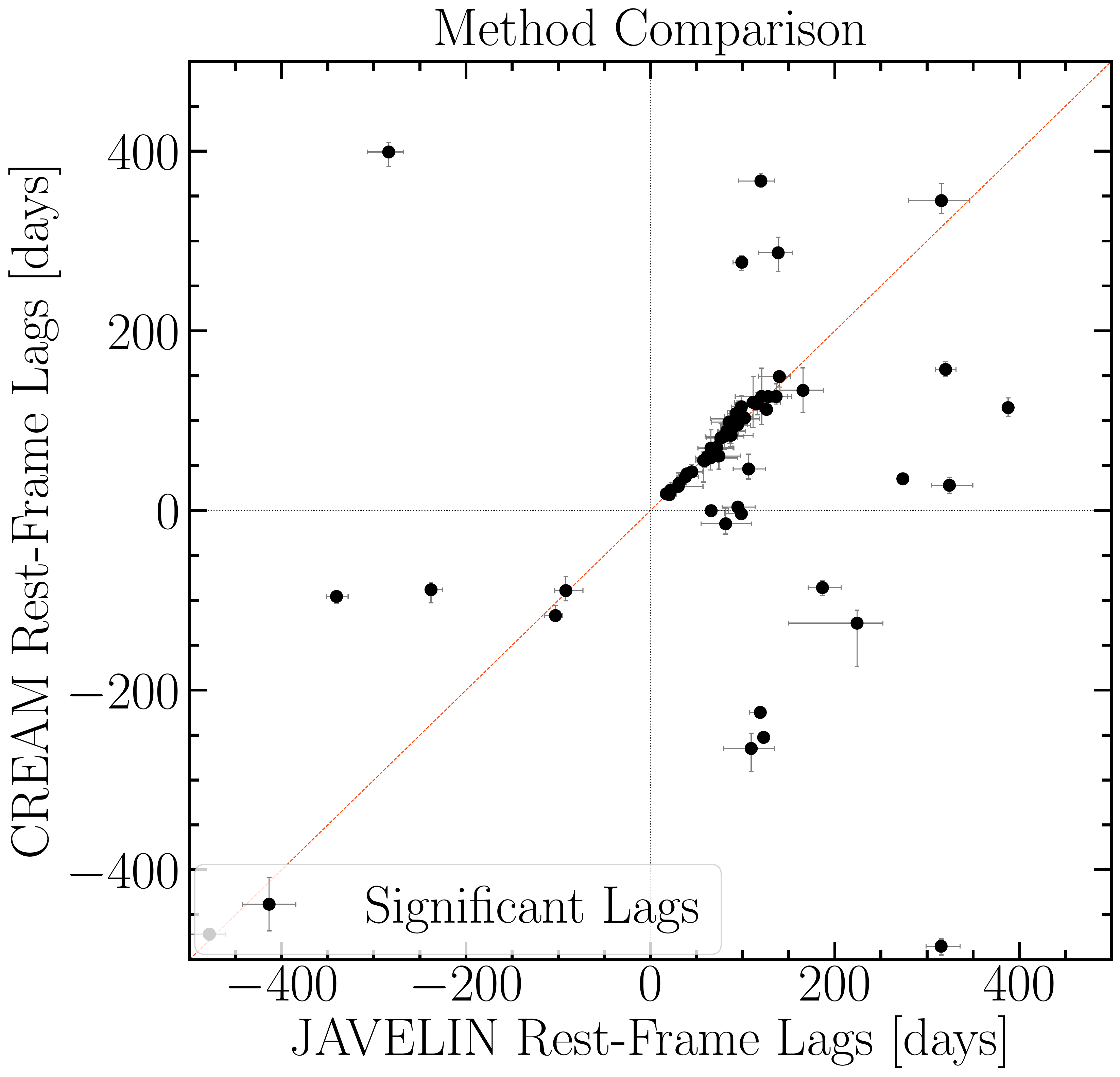}
\caption{A comparison of lag measurements from \jav\ and \cream\ for the sample of 63 positive and negative lags that meet our significance criteria (defined in Section 4.2). Overall, \cream\ and \jav\ lag measurements are consistent within 1$\sigma$ for 39 of the 63 significant lags (62\%), although 33\% of the lag solutions are outliers that differ by more than 3$\sigma$. In many of these outlier cases the \cream\ model fits find lags of $\sim$0 and/or with multiple peaks in the lag posterior that do not meet our significance criteria; only 13 lags are significant in both \jav\ and \cream\ and differ by more than 3$\sigma$.}
\label{fig:lagcompare}
\end{figure}

We test the reliability of our lag detections by comparing the results of \jav\ and \cream, as shown in Figure~\ref{fig:lagcompare}. In general the two methods agree quite well: 60\% of the significant \jav\ lags have \cream\ lags that agree within 1$\sigma$. In the full sample of significant positive and negative lags, there are a large number of outliers (21/63) that have \jav\ and \cream\ lags that differ by more than 3$\sigma$.

\begin{figure*}[t]
\plottwo{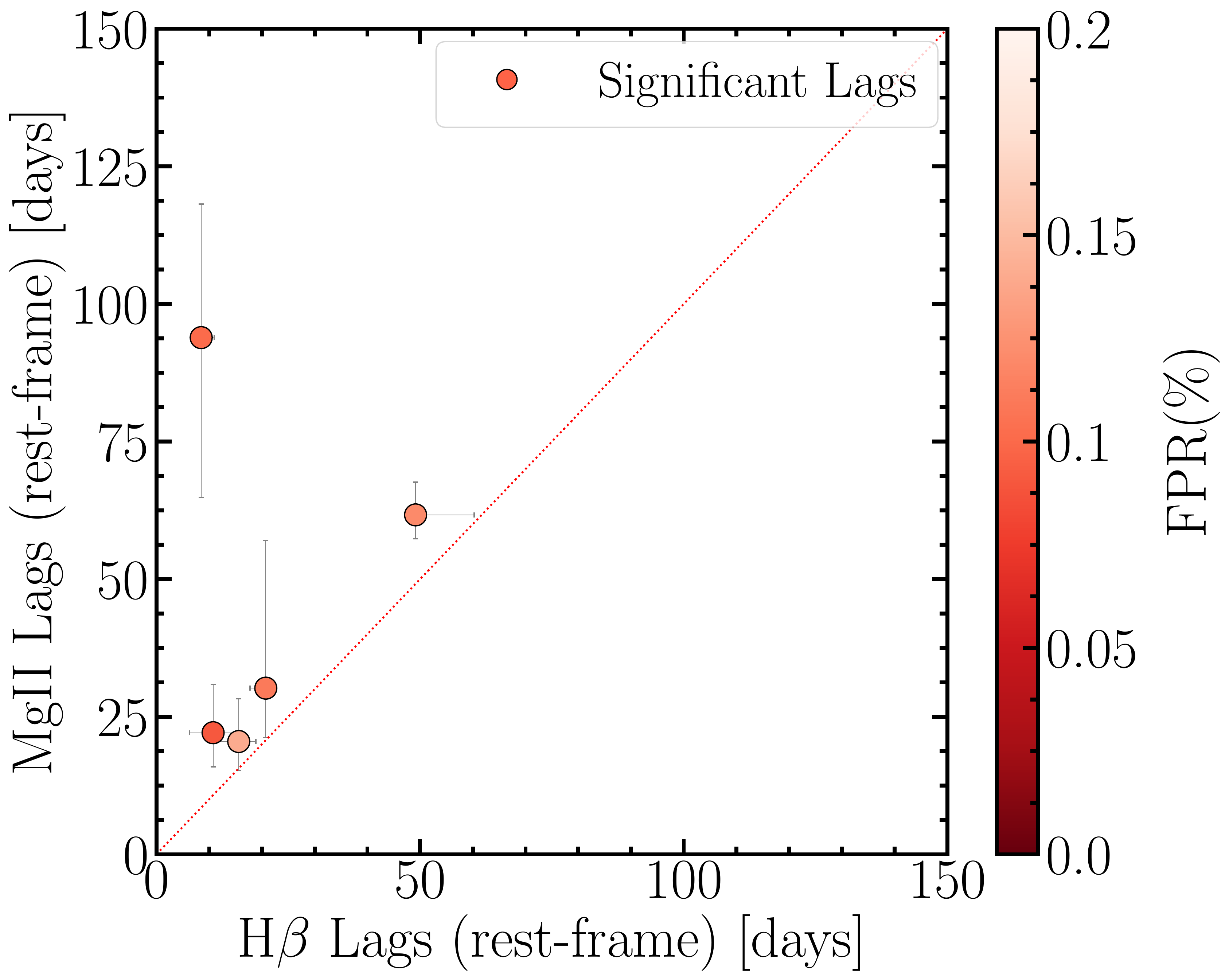}{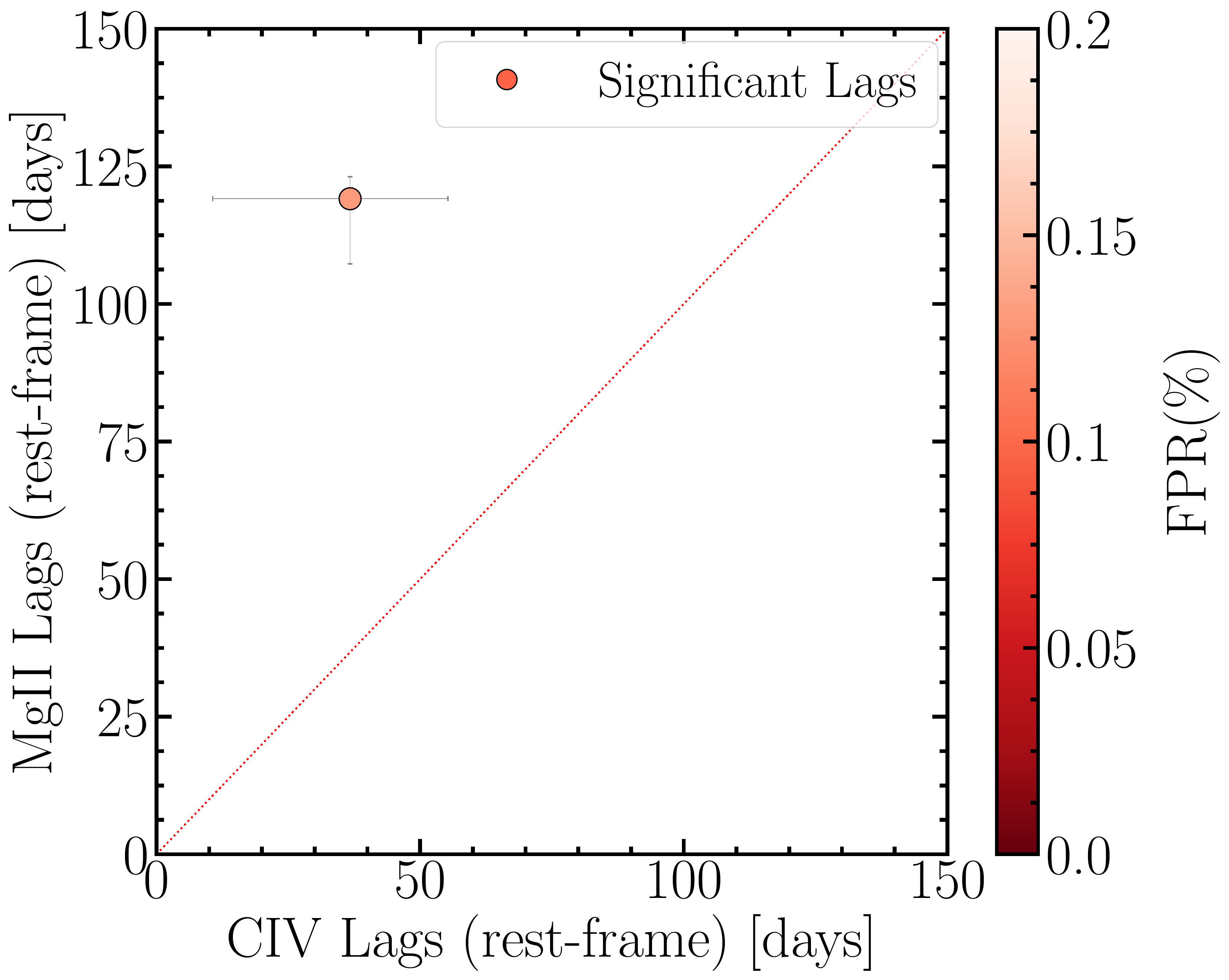}
\caption{The comparison of 5 \MgII\ and \Hb\ lags (\textit{left}) and 1 \MgII\ and \CIV\ lags (\textit{right}) for the quasars with significant lags from both this work and previous SDSS-RM studies \citep{Grier2017,Grier2019}. Limiting the comparison of \MgII\ and \Hb\ lags to the 4 objects with \MgII\ detectable within the 100-day search range of \citet{Grier2017}, the ratio of \MgII\ to \Hb\ lags is $1.4 \pm 0.4$. The single quasar in the right panel has a ratio of \MgII\ to \CIV\ lag ratio of $3.2 \pm 0.6$. In both cases these comparisons are consistent with a stratified BLR, with the \MgII\ emission region at significantly larger radii than \CIV\ and at similar or marginally larger radii than \Hb.}
\label{hbeta_c4_mg2}
\end{figure*}

Visual inspection of the \jav\ and \cream\ model fits leads us to conclude that the \jav\ results are more reliable. In many (8 out of 21) of the outlier cases where the lags disagree by more than 3$\sigma$, the \cream\ lag fit fails to find a significant lag, with a lag posterior centered at $\tau \sim 0$ and/or with multiple peaks and $f_{\rm peak}<60$\%. Recent work by \citet{Li2019} using simulated lightcurves similarly shows that \jav\ typically outperforms other methods of lag identification, with more reliable lag uncertainties and lower false lag detections, for survey-quality RM observations.

We also compare our lag measurements with the 6 \MgII\ lags measured using only the 2014 SDSS-RM data by \citet{Shen2016a}.
We only recover the same lag for 1 of these 6 lags as a positive significant lag (RM\,457). We find a consistent lag with \citet{Shen2016a} for 2 of the 6 (RM\,101 and RM\,229), but the lags do not meet our significance criteria because they have $f_{\rm peak}<0.6$. This is not surprising because \citet{Shen2016a} did not use a $f_{\rm peak}$ criterion for measuring lags. The remaining 3 objects (RM\,589, RM\,767 and RM\,789) are more unusual: the 2014 lightcurves appear to be variable with \MgII\ reverberation, but the other three years have less variability and/or less apparent connection between the \MgII\ and continuum lightcurves, which result in the non-detection of a \MgII\ lag using the 4-year data. These may be examples of anomalous BLR variability, sometimes referred to as ``holiday states'' \citep{Dehghanian2019,Kriss2019}, where the emission line stops reverberating with respect to the optical continuum. 

%%%%%%%%%%%%%%%%%%%%%%%%%%%%%%
%        Discussion          %
%%%%%%%%%%%%%%%%%%%%%%%%%%%%%%
\section{Discussion}
%------------------------------------------------------------------
\subsection{Stratification of the Broad Line Region}\label{sec:blr}
%------------------------------------------------------------------
Reverberation mapping of multiple emission lines can reveal stratification of the broad-line region. Previous work has generally found that high-ionization lines like \CIV\ and \HeII\ generally have shorter lags (i.e., lie closer to the ionizing continuum) while low-ionization lines like \Hb\ and \Ha\ have longer lags \citep[e.g.,][]{Clavel1991, Peterson1999, DeRosa2015}. However, while its lower ionization suggests it is more likely to be emitted at larger radii, it is not clear how \MgII\ fits into the picture of BLR stratification. Unlike the recombination-dominated Balmer lines, the \MgII\ line includes significant collisional excitation, and is expected to have lower responsivity and a broader response function \citep{Goad1993, O'Brien1995, Korista2000, Guo2020}. To date, there have been too few observations of \MgII\ lags to conclusively understand where the \MgII\ line sits relative to the rest of the BLR.

We compare our \MgII\ lags to published SDSS-RM \Hb\ \citep{Grier2017} and \CIV\ \citep{Grier2019} lags in the same quasars in Figure \ref{hbeta_c4_mg2}. There are 7 quasars with both \Hb\ and \MgII\ lags and only 1 quasar with both \CIV\ and \MgII\ lags. The small number of matches is due in part to the limited redshift range for observing both lines; having both \CIV\ and \MgII\ is especially limited because we restricted the \MgII\ sample to $z<1.7$ to avoid variable sky line contamination. The \Hb-\MgII\ lag comparison is further limited by the 100-day search range of the \citet{Grier2017} \Hb\ lag sample, since it excludes longer \Hb\ lags that could be observed in quasars with longer \MgII\ lags.

To avoid this bias, we analysed the 4-year SDSS-RM lightcurves with \jav\ to estimated  \Hb\ lags for the three quasars with \MgII\ lags of $>\,$75~days.
In one of these cases we find the same lag as \citet{Grier2017}, while the other two targets have $f_{\rm peak}<60$\% and the \citet{Grier2017} lags are coincident with secondary peaks in the lag posterior.
 The secondary lag peaks are likely due to additional variability features present in the multi-year data.

Furthermore, the measured lag may be different if the quasar luminosity changed significantly over multiple years of observations. We remove the two sources with low-$f_{\rm peak}$ lags from the comparison and find a \MgII\ to \Hb\ lag ratio of $1.4 \pm 0.4$ (mean and uncertainty in the mean) %1.41 $\pm$ 0.37
for the remaining five objects.
This ratio is consistent with the \MgII\ emitting region being similar in size or marginally larger than the \Hb\ emission region, and is also broadly consistent with previous \MgII\ lag measurements \citep{Clavel1991,Czerny2019}.
A full analysis of the \Hb\ lags measured from the multi-year SDSS-RM data and their comparison to the \MgII\ lags measured here will appear in future work.

The single quasar with both \MgII\ and \CIV\ lags, RM158, has a \MgII\ lag that is $3.2 \pm 0.6$ %$3.24 \pm 0.61$
times longer than the \CIV\ lag. The larger \CIV\ lag is consistent with the BLR stratification model, where high-ionization lines such as \CIV\ are at smaller radii compared to the low-ionization \MgII\ and \Hb\ lines.

%----------------------------------------------------------------
\subsection{The \MgII\ Radius--Luminosity Relation}\label{sec:RL}
%----------------------------------------------------------------

Previous RM studies of \Hb\ and \CIV\ %line measurements 
have established empirical relations between the broad-line lags
and the quasar continuum luminosity \citep{Peterson2005,Kaspi2007,Bentz2013,Du2016a,Grier2017,Lira2018, Hoormann2019, Grier2019}.
These ``radius--luminosity'' relations have typically found a best-fit $R_{\rm{BLR}}\propto \lambda L^{\alpha}_{\lambda}$ consistent with a slope of $\alpha=0.5$, as expected for a photoionization-driven BLR
\citep{Davidson1972}.

In contrast to \Hb\ and \CIV, there has not yet been a sufficient number of \MgII\ lag measurements to construct a \MgII\ $R-L$ relation. Compared to \Hb\ and \CIV, attempts to measure RM \MgII\ lags have been affected by the smaller-amplitude variability of \MgII\ and its slower response to the continuum compared to the Balmer lines (i.e. \citealp{Trevese2007,Woo2008,Hryniewicz2014, Cackett2015}). So far,
there are only $\sim$10 quasars with \MgII\ lag measurements \citep{Clavel1991,Metzroth2006,Lira2018,Czerny2019}, 6 of which come from the 2014 SDSS-RM observations \citep{Shen2016a}. \citet{Czerny2019} combine all the \MgII\ lag measurements from the literature and show that they are broadly consistent with the \Hb\ radius--luminosity relation measured by \citet{Bentz2013} with a slope of $\alpha=0.5$ and a \MgII\ broad-line size similar to \Hb.

We combine our new lag measurements with the existing \MgII\ lag measurements to fit a $R-L$ relation
\begin{equation}
    \log \left(\frac{R_{\rm BLR}}{\rm lt-days}\right) = \beta + \alpha \log \left(\frac{\lum}{10^{44} {\rm erg~s}^{-1}}\right).
\end{equation}
To determine the best-fit $R-L$ relation, we use the PyMc3 GLM robust linear regression method,\footnote{https://docs.pymc.io/notebooks/GLM-robust.html} which takes a Bayesian approach to linear regression.
We includes an intrinsic scatter, $\sigma$, as a fitted parameter added in quadrature to the observed error. This is similar to the intrinsic scatter model used in the \texttt{FITEXY}\footnote{https://github.com/jmeyers314/linmix} method of \citet{Kelly2007}.
% commonly used in reporting the scatter in $R-L$ relation\footnote{We found that PyMc3 is more efficient at sampling compared to \texttt{FITEXY}; this is because our data does not include any non-detection and also the uncertainties on the x-axis is comparably small, leading to a much simpler approach.}.We have investigated that our statistical approach using PyMc3 GLM robust linear regression is in fact similar to the likelihood defined in \citet{Kelly2007}, Equation~20.

\begin{figure}[t]
\centering
\includegraphics[width=90mm]{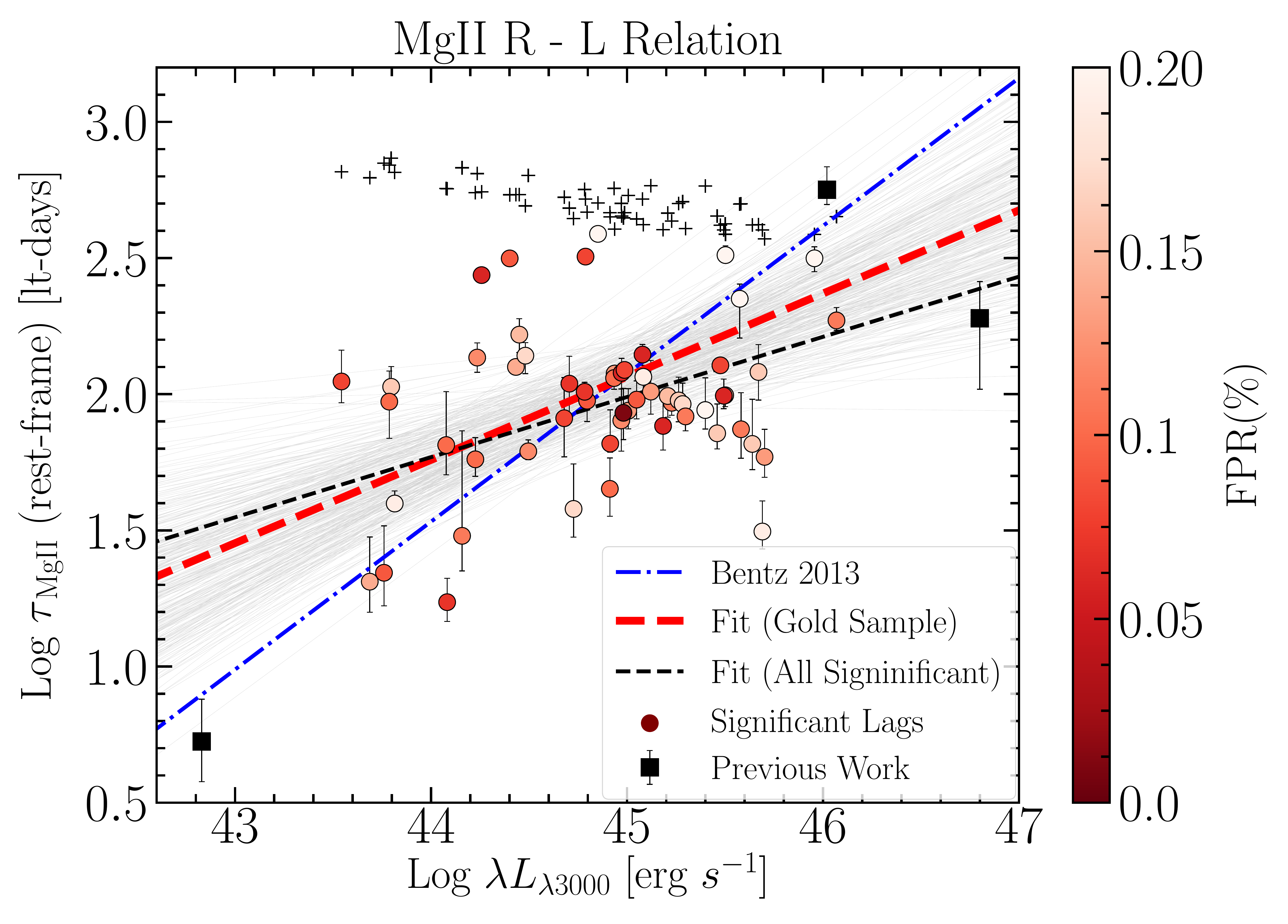}
\caption{The \MgII\ $R-L$ relation for our new \MgII\ lags (circles, color-coded by individual false-positive rate) and previous measurements (black squares, compiled by \citealt{Czerny2019}). The black cross symbols represent the upper limit in rest-frame lag computed from the observed-frame 1000-day search range and the target's redshift. The best-fit linear regression to the previous lags and our ``gold sample'' of lags (with individual false-positive rates of $\leq$10\%) is shown by the dashed red line and has a slope of  0.31$\pm$ 0.1, with an excess intrinsic scatter of 0.36~dex. The gray shading indicates several samples of the MCMC fits. The best-fit line is shallower but marginally consistent (within $2\,\sigma$) with the $\alpha=0.533^{+0.035}_{-0.033}$ slope of the \citet{Bentz2013} best-fit \Hb\ $R-L$ relation, although the \MgII\ $R-L$ relation has significantly larger scatter. Fitting the entire sample, our 57 significant positive lags results in a shallower $R-L$ relation with a slope of $0.22^{+0.06}_{-0.05}$, although this fit is likely affected by a larger number of false-positive lags than the gold sample.
}
\label{fig:radlum}
\end{figure}

Figure \ref{fig:radlum} shows the \MgII\ radius--luminosity relation for our new measurements and the 3 previous \MgII\ lags (compiled by \citealt{Czerny2019}). We use the 24 quasars from the gold sample along with the 3 existing \MgII\ lag measurements to find a best-fit \MgII\ radius--luminosity relation with a slope of $\alpha=0.31^{+0.09}_{-0.10}$ and an intrinsic scatter of 0.36~dex (shown as the red line and gray envelope in Figure~\ref{fig:radlum}), the \MgII\ $R-L$ best-fit slope is shallower but still marginally consistent (within $2\sigma$) with the \Hb\ $R-L$ best-fit line from \citet{Bentz2013}, which lies 
within the uncertainties of our best-fit line in Figure~\ref{fig:radlum}. If we use the $F$-test to quantify whether the slope $\alpha$ is necessary to model the data, we find that a luminosity-independent model ($\alpha\equiv0$) is rejected with a null probability of p = 0.002.
This suggests that there exists a $R-L$ relation for the \MgII\ emission line that is similar to \Hb, as expected for the basic photoionization expectation given the similar ionization potentials of \Hb\ (13.6 eV) and \MgII\ (15.0 eV). The radius-luminosity fit to all 57 significant positive lags has a shallower slope of $0.22^{+0.06}_{-0.05}$, but is likely affected by a larger number of false-positive lags.

The shallower slope of our \MgII\ $R-L$ relation is similar to the shorter \Hb\ lags in SEAMBH and SDSS-RM quasars \citep{Du2016a,Grier2017} compared to the \citet{Bentz2013} relation. As observed for the \Hb\ lags, the shallower best-fit slope may be caused by a range of quasar accretion rates and/or ionization conditions causing shorter \MgII\ lags \citep{Du2019,FonsecaAlvarez2019}. It is likely that that shallower slope of the \MgII\ radius--luminosity relation is connected to its large intrinsic scatter, since large intrinsic scatter tends to lead to a shallower best-fit slope \citep[e.g.][]{Shen2010}. On the other hand, Figure \ref{fig:radlum} shows that the upper limits in rest-frame lag detection (black crosses) are unlikely to affect the measured slope. %\textbf{We also perform a F-test to determine if the best-fit $R-L$ line for the gold sample is a better description of the data than a line of zero slope. The F-test finds a null-probability of $p=0.002$, indicating that the null hypothesis of zero slope is a significantly worse fit to the data than our best-fit $R-L$ relation.}

The best-fit \MgII\ $R-L$ relation has a large excess scatter of 0.36 dex, significantly larger (by $>\,$2$\sigma$) than the 0.25~dex excess scatter measured for the SDSS-RM \Hb\ lags \citep{FonsecaAlvarez2019}. This may be the result of the \MgII\ line having a significant collisional excitation component and/or a broader radial extent in the BLR \citep{Goad1993,Korista2000}. \MgII\ is also a resonance line, so there could be radiative transfer effects that do not occur for \Hb\ line. A broader \MgII\ $R-L$ relation than \Hb\ is also consistent with the predictions of the LOC photoionization models of \citet{Guo2020} which shows that the \MgII\ emitting region is often located where the BLR is truncated and hence less affected by the continuum luminosity. 

It would be interesting to investigate whether the lag offset from the \citet{Bentz2013} relation is connected to the Eddington ratio. Similar studies of the \Hb\ radius--luminosity relation demonstrate that quasars with higher Eddington ratio and/or higher ionization have shorter lags compared to the canonical $R-L$ expectation
\citep{Du2016b, FonsecaAlvarez2019}. 
However, we note that Eddington ratio self-correlates with both axes of the $R-L$ relation and thus is not an independent quantity. A more suitable approach would be to adopt the relative iron strength as a proxy for Eddington ratio \citep[e.g.,][]{Shen2014}. Optical \FeII\ strengths are unavailable in the SDSS spectra of most of our \MgII\ quasars, given their high redshifts, but \citet{Martinez2020} instead found a relationship between relative UV \FeII\ strengths and $R-L$ offset. We plan to further investigate how the \MgII\ $R-L$ relation correlates with other quasar properties in future work.

Our new \MgII\ lag measurements occupy a convenient range of lags between the previous measurements of short lags in nearby low-luminosity Seyfert 1 AGN \citep{Clavel1991, Metzroth2006} and the long lags measured for luminous quasars \citep{Lira2018, Czerny2019}. Future monitoring of the SDSS-RM field with SDSS-V \citep{Kollmeier2019} will cover a 10-year monitoring baseline and add a larger number of longer lags from more luminous quasars.

\begin{table}
\begin{center}
  %{Table 1: \MgII\ $R-L$ Best Fit \label{tbl:bestfit}}
  \begin{tabular}{cccc}
    \hline
    \hline
    Lag Sample & $\alpha$ & $\beta$ & Intrinsic Scatter \\
    \hline
    Significant  & $0.22^{+0.06}_{-0.05}$ & $-7.95^{+5.52}_{-10.45}$ & $0.30^{0.03}_{0.03}$ \\
    Gold  & $0.31^{+0.10}_{-0.10}$ & $-11.69^{+7.34}_{-16.07}$ & $0.36^{+0.07}_{-0.05}$ \\
    \hline
 \end{tabular}
 \caption{\MgII\ $R-L$ Best Fit}
\end{center}
\end{table}

%---------------------------------------------------------------
\subsection{Black Hole Masses at Cosmic High Noon}\label{sec:SE}
%---------------------------------------------------------------

Over the last three decades, numerous campaigns have produced about 100 BH mass measurements from \Hb\ RM of broad-line AGN at $z<0.3$ (e.g., the compilation of \citealt{Bentz2015}). Recent multi-object surveys like SDSS-RM have doubled this number, expanding the sample of \Hb\ RM masses to $z\sim1$ \citep{Shen2016a,Grier2017} and adding a large set of \CIV\ RM masses at $z \sim 2$ \citep{Grier2019}. But there still remains a large gap in RM mass measurements at $1<z<1.5$, where \MgII\ is the only strong broad line available in an observed-frame optical spectrum. This redshift range is particularly important because the peak of SMBH total mass growth occurs within $1<z<2$ \citep[e.g.][]{Aird2015}. 

With so few RM masses available, the bulk of BH masses over cosmic time have been estimated using scaling relations based on the observed \Hb\ $R-L$ relation, substituting a single-epoch luminosity measurement for the expensive RM $R_{\rm BLR}$. Since the $R-L$ relation is only well-measured for \Hb,\ single-epoch masses applying it to \MgII\ and \CIV\ requires an additional scaling from \Hb\ line widths in quasars with both lines \citep{McLure2002, Vestergaard2009, Shen2011, Trakhtenbot2012, Bahk2019}.
Even without this additional step, the uncertainty in \Hb\ single-epoch BH masses is at least 0.4~dex \citep{Vestergaard2006,Shen2013}.
The recently observed $R-L$ offsets of \Hb\ lag measurements in more diverse AGN samples adds additional doubt that the \Hb\ $R-L$ calibrated from previous RM samples describes the broader AGN population \citep{Du2016a,FonsecaAlvarez2019}. Finally, the previous lack of empirical data on the \MgII\ $R-L$ relation raises the question of whether SE masses calibrated for \Hb\ are
reliable for application to \MgII.

\begin{figure}[t]
\centering
\includegraphics[width=85mm]{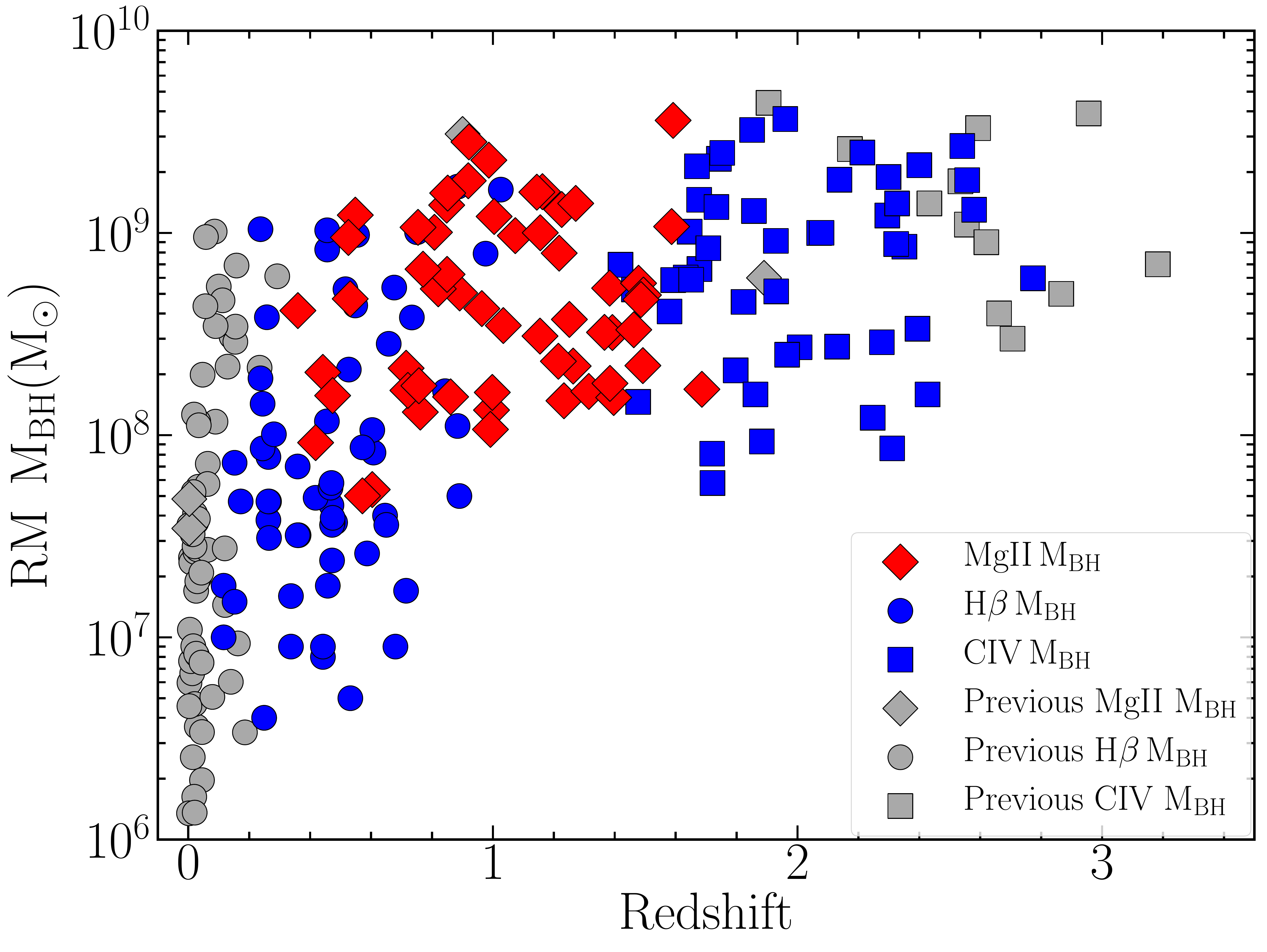}
\caption{RM \Mbh\ vs. redshift for AGNs with RM measurements. Different colored symbols represent the SDSS-RM \Mbh\ measurements from \Hb\ \citep{Grier2017}, \CIV\ \citep{Grier2019} and \MgII\ (this work). Gray symbols illustrate the previous RM \Mbh\ from the same emission lines \citep{Kaspi2007, Bentz2015, Du2016a, Lira2018, Hoormann2019, Czerny2019}.}  
\label{fig:mbh_z}
\end{figure}

We compute RM-based BH masses for the 57 quasars with significant positive \MgII\ lags following Equation \ref{eq:1}. Figure~\ref{fig:mbh_z} shows the new \MgII\ mass measurements with previous RM \Mbh\ from the SDSS-RM and other RM surveys. We use the \MgII\ $\sigma_{line,rms}$ from PrepSpec for the line width, $\Delta V$, and a virial factor $f=4.47$ from \citet{Woo2015}.  We follow the same approach as \citet{Grier2019} and compute the \Mbh\ uncertainties by adding in quadrature the propagated lag and line width errors with an additional 0.16~dex uncertainty, representing the typical uncertainty of RM-based masses from the uncertain $f$-factor \citep{Fausnaugh2016}. The measured masses span
$7.7<\log(\Mbh/M_{\odot})<9.6$ and are included in Table~2.

\begin{figure}[t]
\centering
\includegraphics[width=90mm]{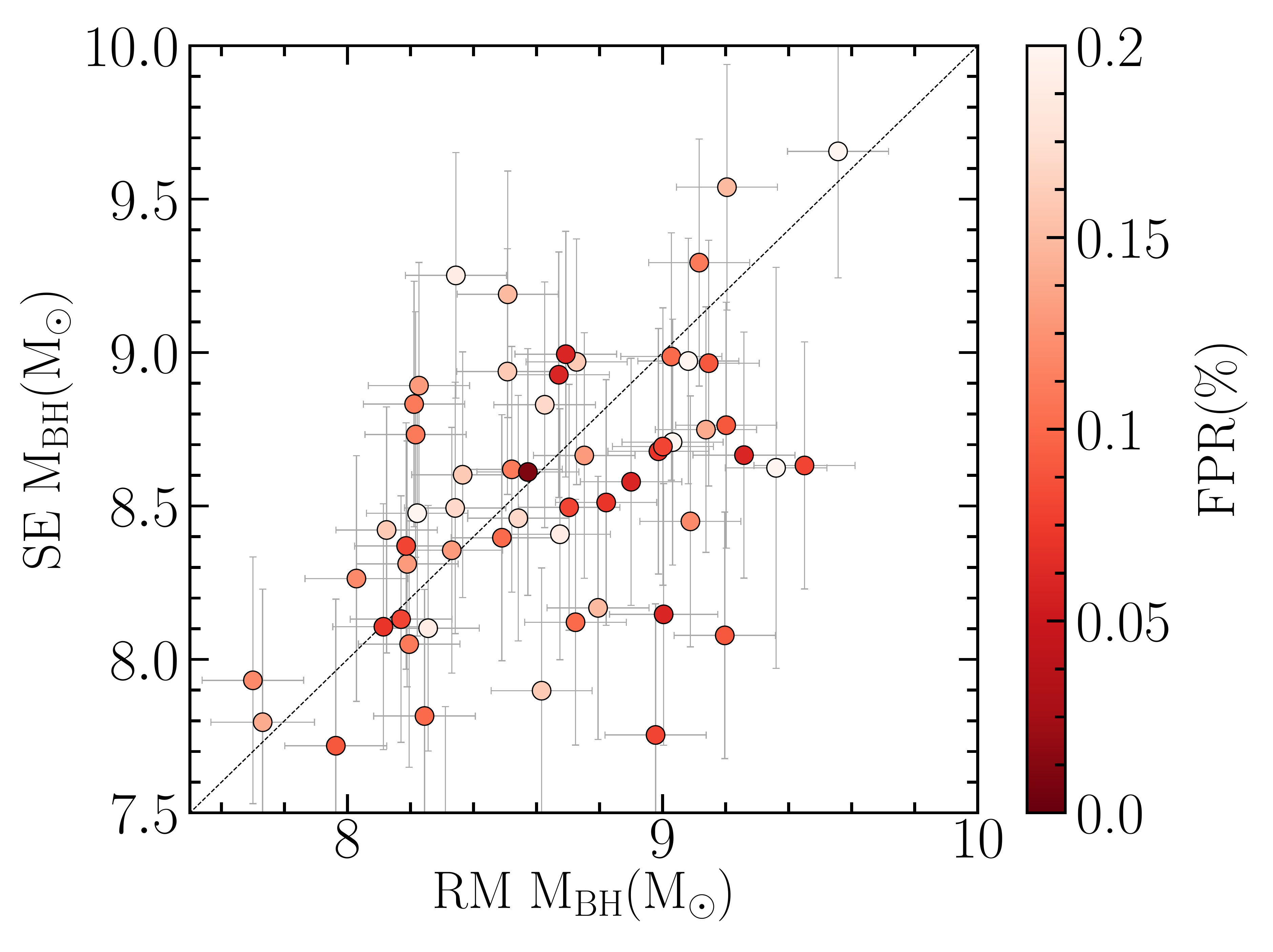}
\caption{Single-epoch \MgII\ \Mbh\ estimates from \citet{Shen2011} compared to the \MgII\ \Mbh\ measurements from RM. The black dashed line shows a 1:1 ratio and measurements are color coded by the individual false positive rate.}  
\label{fig:SE}
\end{figure}

Figure \ref{fig:SE} compares the RM masses with single-epoch masses computed from the \citet{Shen2011} prescription (available in the SDSS-RM sample characterization catalog; \citealt{Shen2019a}).
The RM and single-epoch masses are consistent within their large uncertainties, with an average ratio of 1.000$\pm$0.003 and an excess scatter of 0.45.
The agreement between RM and single-epoch masses is somewhat surprising given the broad scatter in the \MgII\ radius--luminosity relation (Figure \ref{fig:radlum}) and the multi-step scaling required to derive the \MgII\ single-epoch masses \citep[e.g.][]{Vestergaard2009}. The agreement indicates that previous single-epoch masses measured from the \MgII\ line may be reasonable mass estimates within their large uncertainties.

%%%%%%%%%%%%%%%%%%%%%%%%%
%       Summary         %
%%%%%%%%%%%%%%%%%%%%%%%%%

\section{Summary}
We have used four years of SDSS-RM spectroscopic and photometric monitoring data to  measure reverberation lags for the \MgII\ broad emission line. Starting from a sample of 193 quasars with well-detected \MgII\ variability (variability ${\rm SNR} > 20$) in the redshift range $0.35<z<1.7$, we use \jav\ to measure significant positive lags in 57 quasars. Comparing the number of positive and negative significant lags suggests an average false-positive rate of 11\% for the 57 lags. We additionally measure an individual false-positive rates for each quasar by performing \jav\ analysis on shuffled continuum and \MgII\ lightcurves from different objects. 
We use these false-positive rates to define a ``gold sample'' of
24 lag measurements with an individual false-positive rate\,$\le$10\% as our most reliable lag measurements. Our major findings are as follows:

\begin{itemize}
    \item The new \MgII\ lags and previous SDSS-RM measurements of \Hb\ and \CIV\ lags \citep{Grier2017,Grier2019} in the same quasars are consistent with a stratified BLR, with \MgII\ lags that are a factor of a few larger than \CIV\ lags and similar to or slightly larger than \Hb\ lags.
    
    \item We find a radius -- luminosity relation for \MgII\ with a best-fit slope that is shallower but marginally consistent (within $2\,\sigma$) with $\alpha=0.5$, and with 0.4~dex of scatter that is significantly larger than the scatter observed in the \Hb\ radius -- luminosity relation. This implies a broader range of \MgII\ radii than observed for \Hb, consistent with BLR excitation models \citep{Goad1993, O'Brien1995, Korista2000, Guo2020}.
    
    \item We compute RM-based BH masses for the 57 significant positive lags using the measured \MgII\ FWHM and find that the single-epoch masses produced by the prescription of \citet{Shen2011} are consistent with the RM masses.
\end{itemize}
    
The lack of \MgII\ RM measurements at the peak of SMBH growth is among the pressing problems in RM measurements. This work provides the first large set of \MgII\ mass measurements that covers the gap between \Hb\ and \CIV\ in optical RM studies.
Future work will further study BLR stratification using the multi-year SDSS-RM data to measure \Hb\ lags on a longer monitoring baseline that is comparable to the \MgII\ lag measurement limits of this work. We will also further investigate the \MgII\ radius--luminosity relation, using simulations \citep{Li2019, FonsecaAlvarez2019} to understand its shallower slope and large scatter.

\software{PrepSpec, ISIS \citep{Alard1998,Alard2000}, CREAM \citep{Starkey2016}, BOSS reduction pipeline \citep{Dawson2016, Blanton2017}, Javelin \citep{Zu2011}, PyceCREAM (https://github.com/dstar-key23/pycecream),
linmix (https://github.com/jmeyers-314/linmix)
PyMc3 (https://docs.pymc.io/notebooks/-GLM-robust.html)}

\acknowledgments
YH, JRT and GFA acknowledge support from NASA grants HST-GO-15650 and 18-2ADAP18-0177 and NSF grant CAREER-1945546. YS, DAS and JL acknowledge support from an Alfred P. Sloan Research Fellowship (YS) and NSF grant AST-1715579. KH and JVHS acknowledge support from STFC grant ST/R000824/1. PH acknowledges support from the Natural Sciences and Engineering Research Council of Canada (NSERC), funding reference number 2017-05983, and from the National Research Council Canada during his sabbatical at NRC Herzberg Astronomy \& Astrophysics. LCH was supported by the National Key R\&D Program of China (2016YFA0400702) and the National Science Foundation of China (11473002, 11721303). CSK acknowledges support from NSF grants AST-1908952 and AST-1814440.

Funding for SDSS-III was provided by the Alfred P. Sloan Foundation, the Participating Institutions, the National Science Foundation, and the U.S. Department of Energy Office of Science. The SDSS-III web site is http://www.sdss3.org/.  SDSS-III was managed by the Astrophysical Research Consortium for the Participating Institutions of the SDSS-III Collaboration including the University of Arizona, the Brazilian Participation Group, Brookhaven National Laboratory, Carnegie Mellon University, University of Florida, the French Participation Group, the German Participation Group, Harvard University, the Instituto de Astrofisica de Canarias, the Michigan State/Notre Dame/JINA Participation Group, Johns Hopkins University, Lawrence Berkeley National Laboratory, Max Planck Institute for Astrophysics, Max Planck Institute for Extraterrestrial Physics, New Mexico State University, New York University, Ohio State University, Pennsylvania State University, University of Portsmouth, Princeton University, the Spanish Participation Group, University of Tokyo, University of Utah, Vanderbilt University, University of Virginia, University of Washington, and Yale University.

Funding for SDSS-IV has been provided by the Alfred P. Sloan Foundation, the U.S. Department of Energy Office of Science, and the Participating Institutions. SDSS acknowledges support and resources from the Center for High-Performance Computing at the University of Utah. The SDSS web site is www.sdss.org. SDSS is managed by the Astrophysical Research Consortium for the Participating Institutions of the SDSS Collaboration including the Brazilian Participation Group, the Carnegie Institution for Science, Carnegie Mellon University, the Chilean Participation Group, the French Participation Group, Harvard-Smithsonian Center for Astrophysics, Instituto de Astrofísica de Canarias, The Johns Hopkins University, Kavli Institute for the Physics and Mathematics of the Universe (IPMU) / University of Tokyo, the Korean Participation Group, Lawrence Berkeley National Laboratory, Leibniz Institut für Astrophysik Potsdam (AIP), Max-Planck-Institut f\"{u}r Astronomie (MPIA Heidelberg), Max-Planck-Institut f\"{u}r Astrophysik (MPA Garching), Max-Planck-Institut f\"{u}r Extraterrestrische Physik (MPE), National Astronomical Observatories of China, New Mexico State University, New York University, University of Notre Dame, Observat\'{o}rio Nacional / MCTI, The Ohio State University, Pennsylvania State University, Shanghai Astronomical Observatory, United Kingdom Participation Group, Universidad Nacional Aut\'{o}noma de M\'{e}xico, University of Arizona, University of Colorado Boulder, University of Oxford, University of Portsmouth, University of Utah, University of Virginia, University of Washington, University of Wisconsin, Vanderbilt University, and Yale University.

We thank the Bok and CFHT Canadian, Chinese, and French TACs for their support. This research uses Bok data obtained through the Telescope Access Program (TAP), which is funded by the National Astronomical Observatories, Chinese Academy of Sciences, and the Special Fund for Astronomy from the Ministry of Finance in China.  This work is based on observations obtained with MegaPrime/MegaCam, a joint project of CFHT and CEA/DAPNIA, at the Canada-France-Hawaii Telescope (CFHT) which is operated by the National Research Council (NRC) of Canada, the Institut National des Sciences de l'Univers of the Centre National de la Recherche Scientifique of France, and the University of Hawaii. The authors wish to recognize and acknowledge the very significant cultural role and reverence that the summit of Maunakea has always had within the indigenous Hawaiian community. The astronomical community is most fortunate to have the opportunity to conduct observations from this mountain.

\startlongtable
\begin{deluxetable*}{ccccccccccccc}
\tabletypesize{\scriptsize}
\tablecaption{\MgII\ Significant Lag Results \label{tab:table2}}

\tablehead{
\colhead{RMID} & \colhead{$\rm{RA}$} & \colhead{$\rm{DEC}$} & \colhead{$z$} & \colhead{$i$-mag} & \colhead{SNR2} & \colhead{$\log\lambda L_{3000}$} & \colhead{$\tau_{\rm JAV}$} & \colhead{$f_{peak}$} & \colhead{FPR} & \colhead{$\tau_{\rm CREAM}$} & \colhead{log\,\Mbh} & \colhead{Gold} \\
\colhead{} & \colhead{deg} & \colhead{deg} & \colhead{} & \colhead{} & \colhead{} & \colhead{log($\rm erg\,s^{-1}$)} & \colhead{(days)} & \colhead{\%} & \colhead{\%} & \colhead{(days)} & \colhead{($M_{\odot})$} & \colhead{flag} \\
\colhead{} & \colhead{} & \colhead{} & \colhead{} & \colhead{} & \colhead{} & \colhead{} & \colhead{Rest-Frame} & \colhead{} & \colhead{} & \colhead{Rest-Frame} & \colhead{} & \colhead{}
}
\startdata
018 & 213.34694 & 53.1762 & 0.848 & 20.21 & 35 & 44.4 & $125.9^{+6.8}_{-7.0}$ & 74 & 14 & $112.5^{+7.5}_{-6.2}$ & $9.14^{+0.16}_{-0.16}$ & 0 \\ 
028 & 213.92953 & 52.84914 & 1.392 & 19.09 & 36 & 45.6 & $65.7^{+24.8}_{-14.2}$ & 71 & 16 & $69.5^{+20.2}_{-15.1}$ & $8.51^{+0.23}_{-0.19}$ & 0 \\ 
038 & 214.14908 & 52.94704 & 1.383 & 18.76 & 35 & 45.7 & $120.7^{+27.9}_{-28.7}$ & 60 & 16 & $127.0^{+31.5}_{-31.1}$ & $8.73^{+0.19}_{-0.19}$ & 0 \\ 
044 & 214.09516 & 53.30677 & 1.233 & 20.56 & 23 & 44.9 & $65.8^{+18.8}_{-4.8}$ & 88 & 8 & $-0.2^{+1.0}_{-1.3}$ & $8.17^{+0.2}_{-0.16}$ & 1 \\ 
102 & 213.47079 & 52.57895 & 0.861 & 19.54 & 31 & 45.0 & $86.9^{+16.2}_{-13.3}$ & 90 & 13 & $88.4^{+17.1}_{-16.8}$ & $8.19^{+0.18}_{-0.17}$ & 0 \\ 
114 & 213.89293 & 53.62056 & 1.226 & 17.73 & 43 & 46.1 & $186.6^{+20.3}_{-15.4}$ & 67 & 11 & $-85.9^{+7.9}_{-8.9}$ & $9.12^{+0.17}_{-0.16}$ & 0 \\ 
118 & 213.55325 & 52.53583 & 0.715 & 19.32 & 30 & 45.1 & $102.2^{+27.0}_{-19.5}$ & 81 & 13 & $-578.8^{+5.1}_{-1.7}$ & $8.33^{+0.2}_{-0.18}$ & 0 \\ 
123 & 214.65772 & 53.17155 & 0.891 & 20.44 & 26 & 44.7 & $81.6^{+28.0}_{-26.6}$ & 95 & 8 & $-14.8^{+17.6}_{-11.5}$ & $8.7^{+0.22}_{-0.21}$ & 1 \\ 
135 & 212.79563 & 52.80433 & 1.315 & 19.86 & 33 & 45.2 & $93.0^{+9.6}_{-9.8}$ & 68 & 11 & $107.9^{+10.7}_{-14.4}$ & $8.22^{+0.17}_{-0.17}$ & 0 \\ 
158 & 214.47802 & 53.54858 & 1.478 & 20.38 & 20 & 44.9 & $119.1^{+4.0}_{-11.8}$ & 90 & 13 & $-224.8^{+6.0}_{-4.1}$ & $8.75^{+0.16}_{-0.17}$ & 0 \\ 
159 & 213.69478 & 52.42325 & 1.587 & 19.45 & 34 & 45.5 & $324.2^{+25.3}_{-19.4}$ & 76 & 24 & $28.1^{+9.0}_{-8.9}$ & $9.03^{+0.16}_{-0.16}$ & 0 \\ 
160 & 212.67189 & 53.31361 & 0.36 & 19.68 & 189 & 43.8 & $106.5^{+18.2}_{-16.6}$ & 94 & 16 & $46.2^{+16.5}_{-11.2}$ & $8.62^{+0.18}_{-0.17}$ & 0 \\ 
170 & 214.52034 & 53.5496 & 1.163 & 20.17 & 30 & 45.2 & $98.5^{+6.7}_{-17.7}$ & 91 & 15 & $-3.6^{+4.5}_{-3.7}$ & $9.2^{+0.16}_{-0.18}$ & 0 \\ 
185 & 214.39977 & 52.5083 & 0.987 & 19.89 & 20 & 44.9 & $387.9^{+3.3}_{-3.0}$ & 95 & 21 & $114.5^{+10.9}_{-9.8}$ & $9.36^{+0.16}_{-0.16}$ & 0 \\ 
191 & 214.18991 & 53.74633 & 0.442 & 20.45 & 24 & 43.8 & $93.9^{+24.3}_{-29.1}$ & 95 & 10 & $102.1^{+15.0}_{-17.4}$ & $8.31^{+0.2}_{-0.21}$ & 1 \\ 
228 & 214.31267 & 52.38687 & 1.264 & 21.25 & 21 & 44.7 & $37.9^{+14.4}_{-9.1}$ & 75 & 17 & $37.5^{+5.7}_{-5.6}$ & $8.34^{+0.23}_{-0.19}$ & 0 \\ 
232 & 214.21357 & 52.34615 & 0.808 & 20.78 & 25 & 44.3 & $273.8^{+5.1}_{-4.1}$ & 76 & 6 & $35.3^{+5.9}_{-5.2}$ & $9.0^{+0.17}_{-0.17}$ & 1 \\ 
240 & 213.58696 & 52.27498 & 0.762 & 20.88 & 34 & 44.1 & $17.2^{+3.5}_{-2.8}$ & 83 & 7 & $18.6^{+1.5}_{-1.8}$ & $8.11^{+0.18}_{-0.18}$ & 1 \\ 
260 & 212.57517 & 52.57946 & 0.995 & 21.64 & 40 & 45.3 & $94.9^{+18.7}_{-17.2}$ & 96 & 16 & $3.8^{+4.4}_{-2.7}$ & $8.12^{+0.18}_{-0.18}$ & 0 \\ 
280 & 214.95499 & 53.53547 & 1.366 & 19.49 & 42 & 45.5 & $99.1^{+3.3}_{-9.5}$ & 60 & 15 & $276.3^{+7.5}_{-9.1}$ & $8.51^{+0.16}_{-0.17}$ & 0 \\ 
285 & 214.21215 & 52.25793 & 1.034 & 21.3 & 22 & 44.5 & $138.5^{+15.2}_{-21.1}$ & 61 & 17 & $286.8^{+17.5}_{-20.8}$ & $8.54^{+0.17}_{-0.17}$ & 0 \\ 
291 & 214.18017 & 52.24328 & 0.532 & 19.82 & 36 & 43.8 & $39.7^{+4.2}_{-2.6}$ & 87 & 19 & $40.8^{+3.3}_{-3.1}$ & $8.67^{+0.17}_{-0.16}$ & 0 \\ 
294 & 213.42134 & 52.20559 & 1.215 & 19.03 & 25 & 45.5 & $71.8^{+17.8}_{-9.5}$ & 64 & 16 & $70.4^{+8.1}_{-6.8}$ & $8.37^{+0.19}_{-0.17}$ & 0 \\ 
301 & 215.04269 & 52.6749 & 0.548 & 19.76 & 58 & 44.2 & $136.3^{+17.0}_{-16.9}$ & 75 & 12 & $127.0^{+14.1}_{-8.6}$ & $9.09^{+0.17}_{-0.17}$ & 0 \\ 
303 & 214.62585 & 52.37013 & 0.821 & 20.88 & 37 & 44.2 & $57.7^{+10.5}_{-8.3}$ & 85 & 10 & $55.7^{+9.3}_{-24.0}$ & $8.72^{+0.18}_{-0.17}$ & 1 \\ 
329 & 214.249 & 53.96852 & 0.721 & 18.11 & 47 & 45.4 & $87.5^{+23.8}_{-14.0}$ & 69 & 20 & $83.8^{+9.4}_{-9.1}$ & $8.22^{+0.2}_{-0.18}$ & 0 \\ 
338 & 214.98177 & 53.66865 & 0.418 & 20.08 & 20 & 43.8 & $22.1^{+8.8}_{-6.2}$ & 92 & 9 & $22.7^{+8.4}_{-5.5}$ & $7.96^{+0.24}_{-0.2}$ & 1 \\ 
419 & 213.00808 & 52.09101 & 1.272 & 20.35 & 21 & 45.0 & $95.5^{+15.2}_{-15.5}$ & 77 & 9 & $104.7^{+17.2}_{-21.1}$ & $9.15^{+0.17}_{-0.18}$ & 1 \\ 
422 & 211.9132 & 52.98075 & 1.074 & 19.72 & 31 & 44.7 & $109.3^{+25.4}_{-29.6}$ & 72 & 7 & $-264.9^{+16.9}_{-25.5}$ & $8.99^{+0.19}_{-0.2}$ & 1 \\ 
440 & 215.53806 & 53.09994 & 0.754 & 19.53 & 37 & 44.9 & $114.6^{+7.4}_{-10.8}$ & 60 & 10 & $118.7^{+6.7}_{-7.2}$ & $9.03^{+0.16}_{-0.17}$ & 1 \\ 
441 & 213.88294 & 51.98514 & 1.397 & 19.35 & 23 & 45.5 & $127.7^{+5.7}_{-7.3}$ & 60 & 8 & $126.8^{+3.2}_{-5.4}$ & $8.19^{+0.16}_{-0.17}$ & 1 \\ 
449 & 214.92398 & 53.93835 & 1.218 & 20.39 & 21 & 45.0 & $119.8^{+14.7}_{-24.4}$ & 68 & 6 & $366.8^{+8.3}_{-6.8}$ & $8.9^{+0.17}_{-0.18}$ & 1 \\ 
457 & 213.57136 & 51.95628 & 0.604 & 20.29 & 29 & 43.7 & $20.5^{+7.7}_{-5.3}$ & 61 & 14 & $17.6^{+7.0}_{-3.3}$ & $7.73^{+0.23}_{-0.2}$ & 0 \\ 
459 & 213.02897 & 54.14092 & 1.156 & 19.95 & 32 & 45.0 & $122.8^{+5.1}_{-5.7}$ & 79 & 8 & $-252.6^{+3.5}_{-3.2}$ & $9.0^{+0.16}_{-0.16}$ & 1 \\ 
469 & 215.27611 & 53.73527 & 1.004 & 18.31 & 38 & 45.6 & $224.1^{+27.9}_{-74.3}$ & 63 & 24 & $-125.4^{+14.4}_{-48.3}$ & $9.08^{+0.17}_{-0.22}$ & 0 \\ 
492 & 212.97555 & 52.0065 & 0.964 & 18.95 & 31 & 45.3 & $92.0^{+16.3}_{-12.7}$ & 85 & 17 & $94.5^{+11.9}_{-8.0}$ & $8.63^{+0.18}_{-0.17}$ & 0 \\ 
493 & 215.16448 & 52.32457 & 1.592 & 18.6 & 25 & 46.0 & $315.6^{+30.7}_{-35.7}$ & 91 & 21 & $344.9^{+18.8}_{-14.4}$ & $9.56^{+0.17}_{-0.17}$ & 0 \\ 
501 & 214.39663 & 51.98855 & 1.155 & 20.81 & 22 & 44.9 & $44.9^{+11.7}_{-10.4}$ & 70 & 10 & $42.9^{+8.6}_{-4.1}$ & $8.49^{+0.2}_{-0.19}$ & 1 \\ 
505 & 213.15791 & 51.95086 & 1.144 & 20.58 & 21 & 44.8 & $94.7^{+10.8}_{-16.7}$ & 74 & 9 & $95.6^{+11.1}_{-13.3}$ & $9.2^{+0.17}_{-0.18}$ & 1 \\ 
522 & 215.1741 & 52.28379 & 1.384 & 20.21 & 23 & 45.1 & $115.8^{+11.3}_{-16.0}$ & 62 & 19 & $119.1^{+12.3}_{-12.5}$ & $8.26^{+0.17}_{-0.17}$ & 0 \\ 
556 & 215.63556 & 52.66056 & 1.494 & 19.42 & 25 & 45.5 & $98.7^{+13.9}_{-10.8}$ & 66 & 6 & $115.5^{+11.9}_{-14.5}$ & $8.69^{+0.17}_{-0.17}$ & 1 \\ 
588 & 215.7673 & 52.77505 & 0.998 & 18.64 & 44 & 45.6 & $74.3^{+23.0}_{-18.2}$ & 70 & 11 & $60.6^{+12.7}_{-14.6}$ & $8.21^{+0.21}_{-0.19}$ & 0 \\ 
593 & 214.09805 & 51.82018 & 0.992 & 19.84 & 25 & 45.0 & $80.1^{+21.4}_{-20.8}$ & 95 & 12 & $82.6^{+16.7}_{-14.8}$ & $8.03^{+0.2}_{-0.2}$ & 0 \\ 
622 & 212.81328 & 51.86916 & 0.572 & 19.55 & 37 & 44.5 & $61.7^{+6.0}_{-4.3}$ & 94 & 12 & $60.0^{+3.9}_{-4.8}$ & $7.7^{+0.17}_{-0.16}$ & 0 \\ 
645 & 215.16582 & 52.06659 & 0.474 & 19.78 & 22 & 44.2 & $30.2^{+26.8}_{-8.9}$ & 92 & 11 & $26.8^{+15.0}_{-4.5}$ & $8.2^{+0.42}_{-0.21}$ & 0 \\ 
649 & 211.47859 & 52.89651 & 0.85 & 20.48 & 24 & 44.5 & $165.5^{+22.2}_{-25.1}$ & 71 & 15 & $133.7^{+25.1}_{-24.3}$ & $8.8^{+0.17}_{-0.17}$ & 0 \\ 
651 & 215.45543 & 52.24106 & 1.486 & 20.19 & 32 & 45.2 & $76.5^{+18.0}_{-15.6}$ & 97 & 6 & $80.9^{+16.0}_{-16.0}$ & $8.67^{+0.19}_{-0.18}$ & 1 \\ 
675 & 212.18248 & 54.13091 & 0.919 & 19.46 & 38 & 45.1 & $139.8^{+12.0}_{-22.6}$ & 92 & 6 & $149.1^{+3.6}_{-11.7}$ & $9.26^{+0.17}_{-0.18}$ & 1 \\ 
678 & 215.26356 & 52.07418 & 1.463 & 19.62 & 24 & 45.3 & $82.9^{+11.9}_{-10.2}$ & 90 & 11 & $88.4^{+11.7}_{-11.6}$ & $8.52^{+0.17}_{-0.17}$ & 0 \\ 
709 & 212.22948 & 51.9759 & 1.251 & 20.29 & 25 & 45.0 & $85.4^{+17.7}_{-19.3}$ & 73 & 1 & $98.4^{+12.7}_{-23.6}$ & $8.57^{+0.19}_{-0.19}$ & 1 \\ 
714 & 215.95717 & 52.65101 & 0.921 & 19.64 & 51 & 44.8 & $320.1^{+11.3}_{-11.2}$ & 74 & 8 & $157.1^{+8.3}_{-7.8}$ & $9.45^{+0.16}_{-0.16}$ & 1 \\ 
756 & 212.34759 & 51.85559 & 0.852 & 20.29 & 28 & 44.4 & $315.3^{+20.5}_{-16.4}$ & 63 & 9 & $-485.1^{+8.0}_{-9.7}$ & $9.2^{+0.16}_{-0.16}$ & 1 \\ 
761 & 216.05386 & 52.65096 & 0.771 & 20.43 & 48 & 44.8 & $102.1^{+8.2}_{-7.4}$ & 64 & 7 & $103.0^{+7.4}_{-6.6}$ & $8.82^{+0.16}_{-0.16}$ & 1 \\ 
771 & 214.01893 & 54.17766 & 1.492 & 18.64 & 42 & 45.7 & $31.3^{+8.1}_{-4.6}$ & 85 & 19 & $30.5^{+4.4}_{-4.8}$ & $8.34^{+0.2}_{-0.17}$ & 0 \\ 
774 & 212.62967 & 52.05463 & 1.686 & 19.34 & 29 & 45.7 & $58.9^{+13.7}_{-10.1}$ & 95 & 13 & $55.5^{+6.2}_{-7.9}$ & $8.23^{+0.19}_{-0.18}$ & 0 \\ 
792 & 214.503 & 53.3433 & 0.526 & 20.64 & 23 & 43.5 & $111.4^{+29.5}_{-20.0}$ & 92 & 8 & $120.3^{+29.2}_{-28.2}$ & $8.98^{+0.2}_{-0.18}$ & 1 \\ 
848 & 215.60674 & 53.57398 & 0.757 & 20.81 & 25 & 44.1 & $65.1^{+29.4}_{-16.3}$ & 78 & 10 & $58.5^{+15.8}_{-13.5}$ & $8.24^{+0.25}_{-0.19}$ & 1 \\ 
\enddata
\end{deluxetable*}

\bibliography{main.bib}

\begin{thebibliography}{}
\expandafter\ifx\csname natexlab\endcsname\relax\def\natexlab#1{#1}\fi
\providecommand{\url}[1]{\href{#1}{#1}}

\bibitem[{{Aird} {et~al.}(2015){Aird}, {Coil}, {Georgakakis}, {Nandra},
  {Barro}, \& {P{\'e}rez-Gonz{\'a}lez}}]{Aird2015}
{Aird}, J., {Coil}, A.~L., {Georgakakis}, A., {et~al.} 2015, \mnras, 451, 1892

\bibitem[{{Alard}(2000)}]{Alard2000}
{Alard}, C. 2000, \aaps, 144, 363

\bibitem[{{Alard} \& {Lupton}(1998)}]{Alard1998}
{Alard}, C., \& {Lupton}, R.~H. 1998, \apj, 503, 325

\bibitem[{{Aune} {et~al.}(2003){Aune}, {Boulade}, {Charlot}, {Abbon},
  {Borgeaud}, {Carton}, {Carty}, {Da Costa}, {Desforge}, {Deschamps},
  {Eppell{\'e}}, {Gallais}, {Gosset}, {Granelli}, {Gros}, {de Kat}, {Loiseau},
  {Ritou}, {Rouss{\'e}}, {Starzynski}, {Vignal}, \& {Vigroux}}]{Aune2003}
{Aune}, S., {Boulade}, O., {Charlot}, X., {et~al.} 2003, in Society of
  Photo-Optical Instrumentation Engineers (SPIE) Conference Series, Vol. 4841,
  \procspie, ed. M.~{Iye} \& A.~F.~M. {Moorwood}, 513--524

\bibitem[{{Bahk} {et~al.}(2019){Bahk}, {Woo}, \& {Park}}]{Bahk2019}
{Bahk}, H., {Woo}, J.-H., \& {Park}, D. 2019, \apj, 875, 50

\bibitem[{{Barth} {et~al.}(2015){Barth}, {Bennert}, {Canalizo}, {Filippenko},
  {Gates}, {Greene}, {Li}, {Malkan}, {Pancoast}, {Sand }, {Stern}, {Treu},
  {Woo}, {Assef}, {Bae}, {Brewer}, {Cenko}, {Clubb}, {Cooper},
  {Diamond-Stanic}, {Hiner}, {H{\"o}nig}, {Hsiao}, {Kand rashoff}, {Lazarova},
  {Nierenberg}, {Rex}, {Silverman}, {Tollerud}, \& {Walsh}}]{Barth2015}
{Barth}, A.~J., {Bennert}, V.~N., {Canalizo}, G., {et~al.} 2015, \apjs, 217, 26

\bibitem[{{Bentz} \& {Katz}(2015)}]{Bentz2015}
{Bentz}, M.~C., \& {Katz}, S. 2015, Publications of the Astronomical Society of
  the Pacific, 127, 67

\bibitem[{{Bentz} {et~al.}(2009){Bentz}, {Walsh}, {Barth}, {Baliber},
  {Bennert}, {Canalizo}, {Filippenko}, {Ganeshalingam}, {Gates}, {Greene},
  {Hidas}, {Hiner}, {Lee}, {Li}, {Malkan}, {Minezaki}, {Sakata}, {Serduke},
  {Silverman}, {Steele}, {Stern}, {Street}, {Thornton}, {Treu}, {Wang}, {Woo},
  \& {Yoshii}}]{Bentz2009}
{Bentz}, M.~C., {Walsh}, J.~L., {Barth}, A.~J., {et~al.} 2009, \apj, 705, 199

\bibitem[{{Bentz} {et~al.}(2010){Bentz}, {Walsh}, {Barth}, {Yoshii}, {Woo},
  {Wang}, {Treu}, {Thornton}, {Street}, {Steele}, {Silverman}, {Serduke},
  {Sakata}, {Minezaki}, {Malkan}, {Li}, {Lee}, {Hiner}, {Hidas}, {Greene},
  {Gates}, {Ganeshalingam}, {Filippenko}, {Canalizo}, {Bennert}, \&
  {Baliber}}]{Bentz2010}
---. 2010, \apj, 716, 993

\bibitem[{{Bentz} {et~al.}(2013){Bentz}, {Denney}, {Grier}, {Barth},
  {Peterson}, {Vestergaard}, {Bennert}, {Canalizo}, {De Rosa}, {Filippenko},
  {Gates}, {Greene}, {Li}, {Malkan}, {Pogge}, {Stern}, {Treu}, \&
  {Woo}}]{Bentz2013}
{Bentz}, M.~C., {Denney}, K.~D., {Grier}, C.~J., {et~al.} 2013, \apj, 767, 149

\bibitem[{{Blandford} \& {McKee}(1982)}]{Blandford1982}
{Blandford}, R.~D., \& {McKee}, C.~F. 1982, \apj, 255, 419

\bibitem[{{Blanton} {et~al.}(2017){Blanton}, {Bershady}, {Abolfathi},
  {Albareti}, {Allende Prieto}, {Almeida}, {Alonso-Garc{\'\i}a}, {Anders},
  {Anderson}, \& {Andrews}}]{Blanton2017}
{Blanton}, M.~R., {Bershady}, M.~A., {Abolfathi}, B., {et~al.} 2017, \aj, 154,
  28

\bibitem[{{Brandt} \& {Alexander}(2015)}]{Brandt2015}
{Brandt}, W.~N., \& {Alexander}, D.~M. 2015, \aapr, 23, 1

\bibitem[{{Cackett} {et~al.}(2015){Cackett}, {G{\"u}ltekin}, {Bentz},
  {Fausnaugh}, {Peterson}, {Troyer}, \& {Vestergaard}}]{Cackett2015}
{Cackett}, E.~M., {G{\"u}ltekin}, K., {Bentz}, M.~C., {et~al.} 2015, \apj, 810,
  86

\bibitem[{{Cackett} \& {Horne}(2006)}]{Cackett2006}
{Cackett}, E.~M., \& {Horne}, K. 2006, \mnras, 365, 1180

\bibitem[{{Cackett} {et~al.}(2007){Cackett}, {Horne}, \&
  {Winkler}}]{Cackett2007}
{Cackett}, E.~M., {Horne}, K., \& {Winkler}, H. 2007, \mnras, 380, 669

\bibitem[{{Clavel} {et~al.}(1991){Clavel}, {Reichert}, {Alloin}, {Crenshaw},
  {Kriss}, {Krolik}, {Malkan}, {Netzer}, {Peterson}, \&
  {Wamsteker}}]{Clavel1991}
{Clavel}, J., {Reichert}, G.~A., {Alloin}, D., {et~al.} 1991, \apj, 366, 64

\bibitem[{{Czerny} {et~al.}(2019){Czerny}, {Olejak}, {Ralowski}, {Kozlowski},
  {Loli Martinez Aldama}, {Zajacek}, {Pych}, {Hryniewicz}, {Pietrzynski}, \&
  {Sobrino Figaredo}}]{Czerny2019}
{Czerny}, B., {Olejak}, A., {Ralowski}, M., {et~al.} 2019, arXiv e-prints,
  arXiv:1901.09757

\bibitem[{{Davidson}(1972)}]{Davidson1972}
{Davidson}, K. 1972, \apj, 171, 213

\bibitem[{{Dawson} {et~al.}(2013){Dawson}, {Schlegel}, {Ahn}, {Anderson},
  {Aubourg}, {Bailey}, {Barkhouser}, {Bautista}, {Beifiori}, \&
  {Berlind}}]{Dawson2013}
{Dawson}, K.~S., {Schlegel}, D.~J., {Ahn}, C.~P., {et~al.} 2013, \aj, 145, 10

\bibitem[{{Dawson} {et~al.}(2016){Dawson}, {Kneib}, {Percival}, {Alam},
  {Albareti}, {Anderson}, {Armengaud}, {Aubourg}, {Bailey}, \&
  {Bautista}}]{Dawson2016}
{Dawson}, K.~S., {Kneib}, J.-P., {Percival}, W.~J., {et~al.} 2016, \aj, 151, 44

\bibitem[{{De Rosa} {et~al.}(2015){De Rosa}, {Peterson}, {Ely}, {Kriss},
  {Crenshaw}, {Horne}, {Korista}, {Netzer}, {Pogge}, {Ar{\'e}valo}, {Barth},
  {Bentz}, {Brandt}, {Breeveld}, {Brewer}, {Dalla Bont{\`a}}, {De
  Lorenzo-C{\'a}ceres}, {Denney}, {Dietrich}, {Edelson}, {Evans}, {Fausnaugh},
  {Gehrels}, {Gelbord}, {Goad}, {Grier}, {Grupe}, {Hall}, {Kaastra}, {Kelly},
  {Kennea}, {Kochanek}, {Lira}, {Mathur}, {McHardy}, {Nousek}, {Pancoast},
  {Papadakis}, {Pei}, {Schimoia}, {Siegel}, {Starkey}, {Treu}, {Uttley},
  {Vaughan}, {Vestergaard}, {Villforth}, {Yan}, {Young}, \& {Zu}}]{DeRosa2015}
{De Rosa}, G., {Peterson}, B.~M., {Ely}, J., {et~al.} 2015, \apj, 806, 128

\bibitem[{{Dehghanian} {et~al.}(2019){Dehghanian}, {Ferland}, {Kriss},
  {Peterson}, {Mathur}, {Mehdipour}, {Guzm{\'a}n}, {Chatzikos}, {van Hoof},
  {Williams}, {Arav}, {Barth}, {Bentz}, {Bisogni}, {Brandt}, {Crenshaw}, {Dalla
  Bont{\`a}}, {De Rosa}, {Fausnaugh}, {Gelbord}, {Goad}, {Gupta}, {Horne},
  {Kaastra}, {Knigge}, {Korista}, {McHardy}, {Pogge}, {Starkey}, \&
  {Vestergaard}}]{Dehghanian2019}
{Dehghanian}, M., {Ferland}, G.~J., {Kriss}, G.~A., {et~al.} 2019, \apj, 877,
  119

\bibitem[{{Denney} {et~al.}(2016{\natexlab{a}}){Denney}, {Horne}, {Brandt},
  {Grier}, {Ho}, {Peterson}, {Trump}, \& {Ge}}]{Denney2016b}
{Denney}, K.~D., {Horne}, K., {Brandt}, W.~N., {et~al.} 2016{\natexlab{a}},
  \apj, 833, 33

\bibitem[{{Denney} {et~al.}(2010){Denney}, {Peterson}, {Pogge}, {Adair},
  {Atlee}, {Au-Yong}, {Bentz}, {Bird}, {Brokofsky}, {Chisholm}, {Comins},
  {Dietrich}, {Doroshenko}, {Eastman}, {Efimov}, {Ewald}, {Ferbey}, {Gaskell},
  {Hedrick}, {Jackson}, {Klimanov}, {Klimek}, {Kruse}, {Lad{\'e}route}, {Lamb},
  {Leighly}, {Minezaki}, {Nazarov}, {Onken}, {Petersen}, {Peterson},
  {Poindexter}, {Sakata}, {Schlesinger}, {Sergeev}, {Skolski}, {Stieglitz},
  {Tobin}, {Unterborn}, {Vestergaard}, {Watkins}, {Watson}, \&
  {Yoshii}}]{Denney2010}
{Denney}, K.~D., {Peterson}, B.~M., {Pogge}, R.~W., {et~al.} 2010, \apj, 721,
  715

\bibitem[{{Denney} {et~al.}(2016{\natexlab{b}}){Denney}, {Horne}, {Shen},
  {Brandt}, {Ho}, {Peterson}, {Richards}, {Trump}, \& {Ge}}]{Denney2016a}
{Denney}, K.~D., {Horne}, K., {Shen}, Y., {et~al.} 2016{\natexlab{b}}, \apjs,
  224, 14

\bibitem[{{Dexter} {et~al.}(2019){Dexter}, {Xin}, {Shen}, {Grier}, {Liu},
  {Gezari}, {McGreer}, {Brand t}, {Hall}, {Horne}, {Simm}, {Merloni}, {Green},
  {Vivek}, {Trump}, {Homayouni}, {Peterson}, {Schneider}, {Kinemuchi}, {Pan},
  \& {Bizyaev}}]{Dexter2019}
{Dexter}, J., {Xin}, S., {Shen}, Y., {et~al.} 2019, arXiv e-prints,
  arXiv:1906.10138

\bibitem[{{Doi} {et~al.}(2010){Doi}, {Tanaka}, {Fukugita}, {Gunn}, {Yasuda},
  {Ivezi{\'c}}, {Brinkmann}, {de Haars}, {Kleinman}, {Krzesinski}, \& {French
  Leger}}]{Doi2010}
{Doi}, M., {Tanaka}, M., {Fukugita}, M., {et~al.} 2010, \aj, 139, 1628

\bibitem[{{Du} \& {Wang}(2019)}]{Du2019}
{Du}, P., \& {Wang}, J.-M. 2019, \apj, 886, 42

\bibitem[{{Du} {et~al.}(2015){Du}, {Hu}, {Lu}, {Huang}, {Cheng}, {Qiu}, {Li},
  {Zhang}, {Fan}, {Bai}, {Bian}, {Yuan}, {Kaspi}, {Ho}, {Netzer}, {Wang}, \&
  {SEAMBH Collaboration}}]{Du2015}
{Du}, P., {Hu}, C., {Lu}, K.-X., {et~al.} 2015, \apj, 806, 22

\bibitem[{{Du} {et~al.}(2016{\natexlab{a}}){Du}, {Lu}, {Zhang}, {Huang},
  {Wang}, {Hu}, {Qiu}, {Li}, {Fan}, {Fang}, {Bai}, {Bian}, {Yuan}, {Ho},
  {Wang}, \& {SEAMBH Collaboration}}]{Du2016a}
{Du}, P., {Lu}, K.-X., {Zhang}, Z.-X., {et~al.} 2016{\natexlab{a}}, \apj, 825,
  126

\bibitem[{{Du} {et~al.}(2016{\natexlab{b}}){Du}, {Lu}, {Hu}, {Qiu}, {Li},
  {Huang}, {Wang}, {Bai}, {Bian}, {Yuan}, {Ho}, {Wang}, \& {SEAMBH
  Collaboration}}]{Du2016b}
{Du}, P., {Lu}, K.-X., {Hu}, C., {et~al.} 2016{\natexlab{b}}, \apj, 820, 27

\bibitem[{{Event Horizon Telescope Collaboration} {et~al.}(2019){Event Horizon
  Telescope Collaboration}, {Akiyama}, {Alberdi}, {Alef}, {Asada}, {Azulay},
  {Baczko}, {Ball}, {Balokovi{\'c}}, {Barrett}, {Bintley}, {Blackburn},
  {Boland}, {Bouman}, {Bower}, {Bremer}, {Brinkerink}, {Brissenden}, {Britzen},
  {Broderick}, {Broguiere}, {Bronzwaer}, {Byun}, {Carlstrom}, {Chael}, {Chan},
  {Chatterjee}, {Chatterjee}, {Chen}, {Chen}, {Cho}, {Christian}, {Conway},
  {Cordes}, {Crew}, {Cui}, {Davelaar}, {De Laurentis}, {Deane}, {Dempsey},
  {Desvignes}, {Dexter}, {Doeleman}, {Eatough}, {Falcke}, {Fish}, {Fomalont},
  {Fraga-Encinas}, {Friberg}, {Fromm}, {G{\'o}mez}, {Galison}, {Gammie},
  {Garc{\'\i}a}, {Gentaz}, {Georgiev}, {Goddi}, {Gold}, {Gu}, {Gurwell},
  {Hada}, {Hecht}, {Hesper}, {Ho}, {Ho}, {Honma}, {Huang}, {Huang}, {Hughes},
  {Ikeda}, {Inoue}, {Issaoun}, {James}, {Jannuzi}, {Janssen}, {Jeter}, {Jiang},
  {Johnson}, {Jorstad}, {Jung}, {Karami}, {Karuppusamy}, {Kawashima},
  {Keating}, {Kettenis}, {Kim}, {Kim}, {Kim}, {Kino}, {Koay}, {Koch}, {Koyama},
  {Kramer}, {Kramer}, {Krichbaum}, {Kuo}, {Lauer}, {Lee}, {Li}, {Li},
  {Lindqvist}, {Liu}, {Liuzzo}, {Lo}, {Lobanov}, {Loinard}, {Lonsdale}, {Lu},
  {MacDonald}, {Mao}, {Markoff}, {Marrone}, {Marscher}, {Mart{\'\i}-Vidal},
  {Matsushita}, {Matthews}, {Medeiros}, {Menten}, {Mizuno}, {Mizuno}, {Moran},
  {Moriyama}, {Moscibrodzka}, {M{\"u}ller}, {Nagai}, {Nagar}, {Nakamura},
  {Narayan}, {Narayanan}, {Natarajan}, {Neri}, {Ni}, {Noutsos}, {Okino},
  {Olivares}, {Oyama}, {{\"O}zel}, {Palumbo}, {Patel}, {Pen}, {Pesce},
  {Pi{\'e}tu}, {Plambeck}, {PopStefanija}, {Porth}, {Prather},
  {Preciado-L{\'o}pez}, {Psaltis}, {Pu}, {Ramakrishnan}, {Rao}, {Rawlings},
  {Raymond}, {Rezzolla}, {Ripperda}, {Roelofs}, {Rogers}, {Ros}, {Rose},
  {Roshanineshat}, {Rottmann}, {Roy}, {Ruszczyk}, {Ryan}, {Rygl},
  {S{\'a}nchez}, {S{\'a}nchez-Arguelles}, {Sasada}, {Savolainen}, {Schloerb},
  {Schuster}, {Shao}, {Shen}, {Small}, {Sohn}, {SooHoo}, {Tazaki}, {Tiede},
  {Tilanus}, {Titus}, {Toma}, {Torne}, {Trent}, {Trippe}, {Tsuda}, {van
  Bemmel}, {van Langevelde}, {van Rossum}, {Wagner}, {Wardle}, {Weintroub},
  {Wex}, {Wharton}, {Wielgus}, {Wong}, {Wu}, {Young}, {Young}, {Younsi},
  {Yuan}, {Yuan}, {Zensus}, {Zhao}, {Zhao}, {Zhu}, {Farah}, {Meyer-Zhao},
  {Michalik}, {Nadolski}, {Nishioka}, {Pradel}, {Primiani}, {Souccar},
  {Vertatschitsch}, \& {Yamaguchi}}]{EHT_VI}
{Event Horizon Telescope Collaboration}, {Akiyama}, K., {Alberdi}, A., {et~al.}
  2019, \apj, 875, L6

\bibitem[{{Fausnaugh} {et~al.}(2016){Fausnaugh}, {Denney}, {Barth}, {Bentz},
  {Bottorff}, {Carini}, {Croxall}, {De Rosa}, {Goad}, {Horne}, {Joner},
  {Kaspi}, {Kim}, {Klimanov}, {Kochanek}, {Leonard}, {Netzer}, {Peterson},
  {Schn{\"u}lle}, {Sergeev}, {Vestergaard}, {Zheng}, {Zu}, {Anderson},
  {Ar{\'e}valo}, {Bazhaw}, {Borman}, {Boroson}, {Brandt}, {Breeveld}, {Brewer},
  {Cackett}, {Crenshaw}, {Dalla Bont{\`a}}, {De Lorenzo-C{\'a}ceres},
  {Dietrich}, {Edelson}, {Efimova}, {Ely}, {Evans}, {Filippenko}, {Flatland},
  {Gehrels}, {Geier}, {Gelbord}, {Gonzalez}, {Gorjian}, {Grier}, {Grupe},
  {Hall}, {Hicks}, {Horenstein}, {Hutchison}, {Im}, {Jensen}, {Jones},
  {Kaastra}, {Kelly}, {Kennea}, {Kim}, {Korista}, {Kriss}, {Lee}, {Lira},
  {MacInnis}, {Manne-Nicholas}, {Mathur}, {McHardy}, {Montouri}, {Musso},
  {Nazarov}, {Norris}, {Nousek}, {Okhmat}, {Pancoast}, {Papadakis}, {Parks},
  {Pei}, {Pogge}, {Pott}, {Rafter}, {Rix}, {Saylor}, {Schimoia}, {Siegel},
  {Spencer}, {Starkey}, {Sung}, {Teems}, {Treu}, {Turner}, {Uttley},
  {Villforth}, {Weiss}, {Woo}, {Yan}, \& {Young}}]{Fausnaugh2016}
{Fausnaugh}, M.~M., {Denney}, K.~D., {Barth}, A.~J., {et~al.} 2016, \apj, 821,
  56

\bibitem[{{Fonseca Alvarez} {et~al.}(2019){Fonseca Alvarez}, {Trump},
  {Homayouni}, {Grier}, {Shen}, {Horne}, {I-Hsiu Li}, {Brandt}, {Ho},
  {Peterson}, \& {Schneider}}]{FonsecaAlvarez2019}
{Fonseca Alvarez}, G., {Trump}, J.~R., {Homayouni}, Y., {et~al.} 2019, arXiv
  e-prints, arXiv:1910.10719

\bibitem[{{Fukugita} {et~al.}(1996){Fukugita}, {Ichikawa}, {Gunn}, {Doi},
  {Shimasaku}, \& {Schneider}}]{Fukugita1996}
{Fukugita}, M., {Ichikawa}, T., {Gunn}, J.~E., {et~al.} 1996, \aj, 111, 1748

\bibitem[{{Gaskell} \& {Peterson}(1987)}]{Gaskell1987}
{Gaskell}, C.~M., \& {Peterson}, B.~M. 1987, \apjs, 65, 1

\bibitem[{{Gaskell} \& {Sparke}(1986)}]{Gaskell1986}
{Gaskell}, C.~M., \& {Sparke}, L.~S. 1986, \apj, 305, 175

\bibitem[{{Goad} {et~al.}(1993){Goad}, {O'Brien}, \& {Gondhalekar}}]{Goad1993}
{Goad}, M.~R., {O'Brien}, P.~T., \& {Gondhalekar}, P.~M. 1993, \mnras, 263, 149

\bibitem[{{Grier} {et~al.}(2012){Grier}, {Peterson}, {Pogge}, {Denney},
  {Bentz}, {Martini}, {Sergeev}, {Kaspi}, {Minezaki}, {Zu}, {Kochanek},
  {Siverd}, {Shappee}, {Stanek}, {Araya Salvo}, {Beatty}, {Bird}, {Bord},
  {Borman}, {Che}, {Chen}, {Cohen}, {Dietrich}, {Doroshenko}, {Drake},
  {Efimov}, {Free}, {Ginsburg}, {Henderson}, {King}, {Koshida}, {Mogren},
  {Molina}, {Mosquera}, {Nazarov}, {Okhmat}, {Pejcha}, {Rafter}, {Shields},
  {Skowron}, {Szczygiel}, {Valluri}, \& {van Saders}}]{Grier2012}
{Grier}, C.~J., {Peterson}, B.~M., {Pogge}, R.~W., {et~al.} 2012, \apj, 755, 60

\bibitem[{{Grier} {et~al.}(2013){Grier}, {Peterson}, {Horne}, {Bentz}, {Pogge},
  {Denney}, {De Rosa}, {Martini}, {Kochanek}, {Zu}, {Shappee}, {Siverd},
  {Beatty}, {Sergeev}, {Kaspi}, {Araya Salvo}, {Bird}, {Bord}, {Borman}, {Che},
  {Chen}, {Cohen}, {Dietrich}, {Doroshenko}, {Efimov}, {Free}, {Ginsburg},
  {Henderson}, {King}, {Mogren}, {Molina}, {Mosquera}, {Nazarov}, {Okhmat},
  {Pejcha}, {Rafter}, {Shields}, {Skowron}, {Szczygiel}, {Valluri}, \& {van
  Saders}}]{Grier2013}
{Grier}, C.~J., {Peterson}, B.~M., {Horne}, K., {et~al.} 2013, \apj, 764, 47

\bibitem[{{Grier} {et~al.}(2016){Grier}, {Brandt}, {Hall}, {Trump}, {Filiz Ak},
  {Anderson}, {Green}, {Schneider}, {Sun}, {Vivek}, {Beatty}, {Brownstein}, \&
  {Roman-Lopes}}]{Grier2016}
{Grier}, C.~J., {Brandt}, W.~N., {Hall}, P.~B., {et~al.} 2016, \apj, 824, 130

\bibitem[{{Grier} {et~al.}(2017){Grier}, {Trump}, {Shen}, {Horne}, {Kinemuchi},
  {McGreer}, {Starkey}, {Brandt}, {Hall}, {Kochanek}, {Chen}, {Denney},
  {Greene}, {Ho}, {Homayouni}, {I-Hsiu Li}, {Pei}, {Peterson}, {Petitjean},
  {Schneider}, {Sun}, {AlSayyad}, {Bizyaev}, {Brinkmann}, {Brownstein},
  {Bundy}, {Dawson}, {Eftekharzadeh}, {Fernandez-Trincado}, {Gao},
  {Hutchinson}, {Jia}, {Jiang}, {Oravetz}, {Pan}, {Paris}, {Ponder}, {Peters},
  {Rogerson}, {Simmons}, {Smith}, \& {Wang}}]{Grier2017}
{Grier}, C.~J., {Trump}, J.~R., {Shen}, Y., {et~al.} 2017, \apj, 851, 21

\bibitem[{{Grier} {et~al.}(2019){Grier}, {Shen}, {Horne}, {Brandt}, {Trump},
  {Hall}, {Kinemuchi}, {Starkey}, {Schneider}, {Ho}, {Homayouni}, {I-Hsiu Li},
  {McGreer}, {Peterson}, {Bizyaev}, {Chen}, {Dawson}, {Eftekharzadeh}, {Guo},
  {Jia}, {Jiang}, {Kneib}, {Li}, {Li}, {Nie}, {Oravetz}, {Oravetz}, {Pan},
  {Petitjean}, {Ponder}, {Rogerson}, {Vivek}, {Zhang}, \& {Zou}}]{Grier2019}
{Grier}, C.~J., {Shen}, Y., {Horne}, K., {et~al.} 2019, \apj, 887, 38

\bibitem[{{G{\"u}ltekin} {et~al.}(2009){G{\"u}ltekin}, {Richstone}, {Gebhardt},
  {Lauer}, {Tremaine}, {Aller}, {Bender}, {Dressler}, {Faber}, {Filippenko},
  {Green}, {Ho}, {Kormendy}, {Magorrian}, {Pinkney}, \&
  {Siopis}}]{Gultekin2009}
{G{\"u}ltekin}, K., {Richstone}, D.~O., {Gebhardt}, K., {et~al.} 2009, \apj,
  698, 198

\bibitem[{{Gunn} {et~al.}(2006){Gunn}, {Siegmund}, {Mannery}, {Owen}, {Hull},
  {Leger}, {Carey}, {Knapp}, {York}, {Boroski}, {Kent}, {Lupton}, {Rockosi},
  {Evans}, {Waddell}, {Anderson}, {Annis}, {Barentine}, {Bartoszek}, {Bastian},
  {Bracker}, {Brewington}, {Briegel}, {Brinkmann}, {Brown}, {Carr},
  {Czarapata}, {Drennan}, {Dombeck}, {Federwitz}, {Gillespie}, {Gonzales},
  {Hansen}, {Harvanek}, {Hayes}, {Jordan}, {Kinney}, {Klaene}, {Kleinman},
  {Kron}, {Kresinski}, {Lee}, {Limmongkol}, {Lindenmeyer}, {Long}, {Loomis},
  {McGehee}, {Mantsch}, {Neilsen}, {Neswold}, {Newman}, {Nitta}, {Peoples},
  {Pier}, {Prieto}, {Prosapio}, {Rivetta}, {Schneider}, {Snedden}, \&
  {Wang}}]{Gunn2006}
{Gunn}, J.~E., {Siegmund}, W.~A., {Mannery}, E.~J., {et~al.} 2006, \aj, 131,
  2332

\bibitem[{{Guo} {et~al.}(2019){Guo}, {Shen}, {He}, {Wang}, {Liu}, {Wang},
  {Sun}, {Yang}, {Kong}, \& {Sheng}}]{Guo2019}
{Guo}, H., {Shen}, Y., {He}, Z., {et~al.} 2019, arXiv e-prints,
  arXiv:1907.06669

\bibitem[{{Guo} {et~al.}(2020){Guo}, {Shen}, {He}, {Wang}, {Liu}, {Wang},
  {Sun}, {Yang}, {Kong}, \& {Sheng}}]{Guo2020}
---. 2020, \apj, 888, 58

\bibitem[{{Hemler} {et~al.}(2019){Hemler}, {Grier}, {Brandt}, {Hall}, {Horne},
  {Shen}, {Trump}, {Schneider}, {Vivek}, {Bizyaev}, {Oravetz}, {Oravetz}, \&
  {Pan}}]{Hemler2019}
{Hemler}, Z.~S., {Grier}, C.~J., {Brandt}, W.~N., {et~al.} 2019, \apj, 872, 21

\bibitem[{{Homayouni} {et~al.}(2019){Homayouni}, {Trump}, {Grier}, {Shen},
  {Starkey}, {Brandt}, {Fonseca Alvarez}, {Hall}, {Horne}, {Kinemuchi}, {I-Hsiu
  Li}, {McGreer}, {Sun}, {Ho}, \& {Schneider}}]{Homayouni2019}
{Homayouni}, Y., {Trump}, J.~R., {Grier}, C.~J., {et~al.} 2019, \apj, 880, 126

\bibitem[{{Hoormann} {et~al.}(2019){Hoormann}, {Martini}, {Davis}, {King},
  {Lidman}, {Mudd}, {Sharp}, {Sommer}, {Tucker}, \& {Yu}}]{Hoormann2019}
{Hoormann}, J.~K., {Martini}, P., {Davis}, T.~M., {et~al.} 2019, \mnras, 487,
  3650

\bibitem[{{Hryniewicz} {et~al.}(2014){Hryniewicz}, {Czerny}, {Pych}, {Udalski},
  {Krupa}, {{\'S}wi{\c{e}}to{\'n}}, \& {Kaluzny}}]{Hryniewicz2014}
{Hryniewicz}, K., {Czerny}, B., {Pych}, W., {et~al.} 2014, \aap, 562, A34

\bibitem[{{Hu} {et~al.}(2015){Hu}, {Du}, {Lu}, {Li}, {Wang}, {Qiu}, {Bai},
  {Kaspi}, {Ho}, {Netzer}, {Wang}, \& {SEAMBH Collaboration}}]{Hu2015}
{Hu}, C., {Du}, P., {Lu}, K.-X., {et~al.} 2015, \apj, 804, 138

\bibitem[{{Kaspi} {et~al.}(2007){Kaspi}, {Brandt}, {Maoz}, {Netzer},
  {Schneider}, \& {Shemmer}}]{Kaspi2007}
{Kaspi}, S., {Brandt}, W.~N., {Maoz}, D., {et~al.} 2007, \apj, 659, 997

\bibitem[{{Kaspi} {et~al.}(2000){Kaspi}, {Smith}, {Netzer}, {Maoz}, {Jannuzi},
  \& {Giveon}}]{Kaspi2000}
{Kaspi}, S., {Smith}, P.~S., {Netzer}, H., {et~al.} 2000, \apj, 533, 631

\bibitem[{{Kelly}(2007)}]{Kelly2007}
{Kelly}, B.~C. 2007, \apj, 665, 1489

\bibitem[{{Kelly} {et~al.}(2009){Kelly}, {Bechtold}, \&
  {Siemiginowska}}]{Kelly2009}
{Kelly}, B.~C., {Bechtold}, J., \& {Siemiginowska}, A. 2009, \apj, 698, 895

\bibitem[{{Kollmeier} {et~al.}(2019){Kollmeier}, {Anderson}, {Blanc},
  {Blanton}, {Covey}, {Crane}, {Drory}, {Frinchaboy}, {Froning}, {Johnson},
  {Kneib}, {Kreckel}, {Merloni}, {Pellegrini}, {Pogge}, {Ramirez}, {Rix},
  {Sayres}, {S{\'a}nchez-Gallego}, {Shen}, {Tkachenko}, {Trump}, {Tuttle},
  {Weijmans}, {Zasowski}, {Barbuy}, {Beaton}, {Bergemann}, {Bochanski},
  {Brandt}, {Casey}, {Cherinka}, {Eracleous}, {Fan}, {Garc{\'\i}a}, {Green},
  {Hekker}, {Lane}, {Longa-Pe{\~n}a}, {Mathur}, {Meza}, {Minchev}, {Myers},
  {Nidever}, {Nitschelm}, {O'Connell}, {Price-Whelan}, {Raddick}, {Rossi},
  {Sankrit}, {Simon}, {Stutz}, {Ting}, {Trakhtenbrot}, {Weaver}, {Willmer}, \&
  {Weinberg}}]{Kollmeier2019}
{Kollmeier}, J., {Anderson}, S.~F., {Blanc}, G.~A., {et~al.} 2019, in \baas,
  Vol.~51, 274

\bibitem[{{Korista} \& {Goad}(2000)}]{Korista2000}
{Korista}, K.~T., \& {Goad}, M.~R. 2000, \apj, 536, 284

\bibitem[{{Kormendy} \& {Ho}(2013)}]{Kormendy2013}
{Kormendy}, J., \& {Ho}, L.~C. 2013, arXiv e-prints, arXiv:1308.6483

\bibitem[{{Koz{\l}owski}(2016)}]{Kozlowski2016}
{Koz{\l}owski}, S. 2016, \apj, 826, 118

\bibitem[{{Kriss} {et~al.}(2019){Kriss}, {De Rosa}, {Ely}, {Peterson},
  {Kaastra}, {Mehdipour}, {Ferland}, {Dehghanian}, {Mathur}, {Edelson},
  {Korista}, {Arav}, {Barth}, {Bentz}, {Brandt}, {Crenshaw}, {Dalla Bont{\`a}},
  {Denney}, {Done}, {Eracleous}, {Fausnaugh}, {Gardner}, {Goad}, {Grier},
  {Horne}, {Kochanek}, {McHardy}, {Netzer}, {Pancoast}, {Pei}, {Pogge},
  {Proga}, {Silva}, {Tejos}, {Vestergaard}, {Adams}, {Anderson}, {Ar{\'e}valo},
  {Beatty}, {Behar}, {Bennert}, {Bianchi}, {Bigley}, {Bisogni},
  {Boissay-Malaquin}, {Borman}, {Bottorff}, {Breeveld}, {Brotherton}, {Brown},
  {Brown}, {Cackett}, {Canalizo}, {Cappi}, {Carini}, {Clubb}, {Comerford},
  {Coker}, {Corsini}, {Costantini}, {Croft}, {Croxall}, {Deason}, {De
  Lorenzo-C{\'a}ceres}, {De Marco}, {Dietrich}, {Di Gesu}, {Ebrero}, {Evans},
  {Filippenko}, {Flatland}, {Gates}, {Gehrels}, {Geier}, {Gelbord}, {Gonzalez},
  {Gorjian}, {Grupe}, {Gupta}, {Hall}, {Henderson}, {Hicks}, {Holmbeck},
  {Holoien}, {Hutchison}, {Im}, {Jensen}, {Johnson}, {Joner}, {Kaspi}, {Kelly},
  {Kelly}, {Kennea}, {Kim}, {Kim}, {Kim}, {King}, {Klimanov}, {Krongold},
  {Lau}, {Lee}, {Leonard}, {Li}, {Lira}, {Lochhaas}, {Ma}, {MacInnis},
  {Malkan}, {Manne-Nicholas}, {Matt}, {Mauerhan}, {McGurk}, {Montuori},
  {Morelli}, {Mosquera}, {Mudd}, {M{\"u}ller-S{\'a}nchez}, {Nazarov}, {Norris},
  {Nousek}, {Nguyen}, {Ochner}, {Okhmat}, {Paltani}, {Parks}, {Pinto},
  {Pizzella}, {Poleski}, {Ponti}, {Pott}, {Rafter}, {Rix}, {Runnoe}, {Saylor},
  {Schimoia}, {Schn{\"u}lle}, {Scott}, {Sergeev}, {Shappee}, {Shivvers},
  {Siegel}, {Simonian}, {Siviero}, {Skielboe}, {Somers}, {Spencer}, {Starkey},
  {Stevens}, {Sung}, {Tayar}, {Teems}, {Treu}, {Turner}, {Uttley}, {. Van
  Saders}, {Vican}, {Villforth}, {Villanueva}, {Walton}, {Waters}, {Weiss},
  {Woo}, {Yan}, {Yuk}, {Zheng}, {Zhu}, \& {Zu}}]{Kriss2019}
{Kriss}, G.~A., {De Rosa}, G., {Ely}, J., {et~al.} 2019, \apj, 881, 153

\bibitem[{{Li} {et~al.}(2019){Li}, {Shen}, {Brandt}, {Grier}, {Hall}, {Ho},
  {Homayouni}, {Horne}, {Schneider}, {Trump}, \& {Starkey}}]{Li2019}
{Li}, J., {Shen}, Y., {Brandt}, W.~N., {et~al.} 2019, \apj, 884, 119

\bibitem[{{Lira} {et~al.}(2018){Lira}, {Kaspi}, {Netzer}, {Botti}, {Morrell},
  {Mej{\'\i}a-Restrepo}, {S{\'a}nchez-S{\'a}ez}, {Mart{\'\i}nez-Palomera}, \&
  {L{\'o}pez}}]{Lira2018}
{Lira}, P., {Kaspi}, S., {Netzer}, H., {et~al.} 2018, \apj, 865, 56

\bibitem[{{MacLeod} {et~al.}(2010){MacLeod}, {Ivezi{\'c}}, {Kochanek},
  {Koz{\l}owski}, {Kelly}, {Bullock}, {Kimball}, {Sesar}, {Westman}, {Brooks},
  {Gibson}, {Becker}, \& {de Vries}}]{MacLeod2010}
{MacLeod}, C.~L., {Ivezi{\'c}}, {\v{Z}}., {Kochanek}, C.~S., {et~al.} 2010,
  \apj, 721, 1014

\bibitem[{{MacLeod} {et~al.}(2012){MacLeod}, {Ivezi{\'c}}, {Sesar}, {de Vries},
  {Kochanek}, {Kelly}, {Becker}, {Lupton}, {Hall}, {Richards}, {Anderson}, \&
  {Schneider}}]{MacLeod2012}
{MacLeod}, C.~L., {Ivezi{\'c}}, {\v{Z}}., {Sesar}, B., {et~al.} 2012, \apj,
  753, 106

\bibitem[{{Magorrian} {et~al.}(1998){Magorrian}, {Tremaine}, {Richstone},
  {Bender}, {Bower}, {Dressler}, {Faber}, {Gebhardt}, {Green}, {Grillmair},
  {Kormendy}, \& {Lauer}}]{Magorrian1998}
{Magorrian}, J., {Tremaine}, S., {Richstone}, D., {et~al.} 1998, \aj, 115, 2285

\bibitem[{{Mart{\'\i}nez-Aldama} {et~al.}(2020){Mart{\'\i}nez-Aldama},
  {Zaja{\^{c}}ek}, {Czerny}, \& {Panda}}]{Martinez2020}
{Mart{\'\i}nez-Aldama}, M.~L., {Zaja{\^{c}}ek}, M., {Czerny}, B., \& {Panda},
  S. 2020, arXiv e-prints, arXiv:2007.09955

\bibitem[{{Matsuoka} {et~al.}(2015){Matsuoka}, {Strauss}, {Shen}, {Brand t},
  {Greene}, {Ho}, {Schneider}, {Sun}, \& {Trump}}]{Matsuoka2015}
{Matsuoka}, Y., {Strauss}, M.~A., {Shen}, Y., {et~al.} 2015, \apj, 811, 91

\bibitem[{{McLure} \& {Jarvis}(2002)}]{McLure2002}
{McLure}, R.~J., \& {Jarvis}, M.~J. 2002, Monthly Notices of the Royal
  Astronomical Society, 337, 109

\bibitem[{{Metzroth} {et~al.}(2006){Metzroth}, {Onken}, \&
  {Peterson}}]{Metzroth2006}
{Metzroth}, K.~G., {Onken}, C.~A., \& {Peterson}, B.~M. 2006, \apj, 647, 901

\bibitem[{{O'Brien} {et~al.}(1995){O'Brien}, {Goad}, \&
  {Gondhalekar}}]{O'Brien1995}
{O'Brien}, P.~T., {Goad}, M.~R., \& {Gondhalekar}, P.~M. 1995, \mnras, 275,
  1125

\bibitem[{{Pancoast} {et~al.}(2018){Pancoast}, {Barth}, {Horne}, {Treu},
  {Brewer}, {Bennert}, {Canalizo}, {Gates}, {Li}, {Malkan}, {Sand}, {Schmidt},
  {Valenti}, {Woo}, {Clubb}, {Cooper}, {Crawford}, {H{\"o}nig}, {Joner},
  {Kandrashoff}, {Lazarova}, {Nierenberg}, {Romero-Colmenero}, {Son},
  {Tollerud}, {Walsh}, \& {Winkler}}]{Pancoast2018}
{Pancoast}, A., {Barth}, A.~J., {Horne}, K., {et~al.} 2018, \apj, 856, 108

\bibitem[{{Pei} {et~al.}(2017){Pei}, {Fausnaugh}, {Barth}, {Peterson}, {Bentz},
  {De Rosa}, {Denney}, {Goad}, {Kochanek}, {Korista}, {Kriss}, {Pogge},
  {Bennert}, {Brotherton}, {Clubb}, {Dalla Bont{\`a}}, {Filippenko}, {Greene},
  {Grier}, {Vestergaard}, {Zheng}, {Adams}, {Beatty}, {Bigley}, {Brown},
  {Brown}, {Canalizo}, {Comerford}, {Coker}, {Corsini}, {Croft}, {Croxall},
  {Deason}, {Eracleous}, {Fox}, {Gates}, {Henderson}, {Holmbeck}, {Holoien},
  {Jensen}, {Johnson}, {Kelly}, {Kim}, {King}, {Lau}, {Li}, {Lochhaas}, {Ma},
  {Manne-Nicholas}, {Mauerhan}, {Malkan}, {McGurk}, {Morelli}, {Mosquera},
  {Mudd}, {Muller Sanchez}, {Nguyen}, {Ochner}, {Ou-Yang}, {Pancoast}, {Penny},
  {Pizzella}, {Poleski}, {Runnoe}, {Scott}, {Schimoia}, {Shappee}, {Shivvers},
  {Simonian}, {Siviero}, {Somers}, {Stevens}, {Strauss}, {Tayar}, {Tejos},
  {Treu}, {Van Saders}, {Vican}, {Villanueva}, {Yuk}, {Zakamska}, {Zhu},
  {Anderson}, {Ar{\'e}valo}, {Bazhaw}, {Bisogni}, {Borman}, {Bottorff},
  {Brandt}, {Breeveld}, {Cackett}, {Carini}, {Crenshaw}, {De
  Lorenzo-C{\'a}ceres}, {Dietrich}, {Edelson}, {Efimova}, {Ely}, {Evans},
  {Ferland}, {Flatland}, {Gehrels}, {Geier}, {Gelbord}, {Grupe}, {Gupta},
  {Hall}, {Hicks}, {Horenstein}, {Horne}, {Hutchison}, {Im}, {Joner}, {Jones},
  {Kaastra}, {Kaspi}, {Kelly}, {Kennea}, {Kim}, {Kim}, {Klimanov}, {Lee},
  {Leonard}, {Lira}, {MacInnis}, {Mathur}, {McHardy}, {Montouri}, {Musso},
  {Nazarov}, {Netzer}, {Norris}, {Nousek}, {Okhmat}, {Papadakis}, {Parks},
  {Pott}, {Rafter}, {Rix}, {Saylor}, {Schn{\"u}lle}, {Sergeev}, {Siegel},
  {Skielboe}, {Spencer}, {Starkey}, {Sung}, {Teems}, {Turner}, {Uttley},
  {Villforth}, {Weiss}, {Woo}, {Yan}, {Young}, \& {Zu}}]{Pei2017}
{Pei}, L., {Fausnaugh}, M.~M., {Barth}, A.~J., {et~al.} 2017, \apj, 837, 131

\bibitem[{{Peterson}(1993)}]{Peterson1993}
{Peterson}, B.~M. 1993, Publications of the Astronomical Society of the
  Pacific, 105, 247

\bibitem[{{Peterson}(2004)}]{Peterson2004}
{Peterson}, B.~M. 2004, in The Interplay Among Black Holes, Stars and ISM in
  Galactic Nuclei, ed. T.~{Storchi-Bergmann}, L.~C. {Ho}, \& H.~R. {Schmitt},
  Vol. 222, 15--20

\bibitem[{{Peterson} \& {Wandel}(1999)}]{Peterson1999}
{Peterson}, B.~M., \& {Wandel}, A. 1999, \apjl, 521, L95

\bibitem[{{Peterson} {et~al.}(1991){Peterson}, {Balonek}, {Barker}, {Bechtold},
  {Bertram}, {Bochkarev}, {Bolte}, {Bond}, {Boroson}, {Carini}, {Carone},
  {Christensen}, {Clements}, {Cochran}, {Cohen}, {Crampton}, {Dietrich},
  {Elvis}, {Ferguson}, {Filippenko}, {Fricke}, {Gaskell}, {Halpern}, {Huchra},
  {Hutchings}, {Kollatschny}, {Koratkar}, {Korista}, {Krolik}, {Lame}, {Laor},
  {Leacock}, {MacAlpine}, {Malkan}, {Maoz}, {Miller}, {Morris}, {Netzer},
  {Oliveira}, {Penfold}, {Penston}, {Perez}, {Pogge}, {Richmond}, {Romanishin},
  {Rosenblatt}, {Saddlemyer}, {Sadun}, {Sawyer}, {Shields}, {Shapovalova},
  {Smith}, {Smith}, {Smith}, {Sun}, {Thiele}, {Turner}, {Veilleux}, {Wagner},
  {Weymann}, {Wilkes}, {Wills}, {Wills}, \& {Younger}}]{Peterson1991}
{Peterson}, B.~M., {Balonek}, T.~J., {Barker}, E.~S., {et~al.} 1991, \apj, 368,
  119

\bibitem[{{Peterson} {et~al.}(2005){Peterson}, {Bentz}, {Desroches},
  {Filippenko}, {Ho}, {Kaspi}, {Laor}, {Maoz}, {Moran}, {Pogge}, \&
  {Quillen}}]{Peterson2005}
{Peterson}, B.~M., {Bentz}, M.~C., {Desroches}, L.-B., {et~al.} 2005, \apj,
  632, 799

\bibitem[{{Reichert} {et~al.}(1994){Reichert}, {Rodriguez-Pascual}, {Alloin},
  {Clavel}, {Crenshaw}, {Kriss}, {Krolik}, {Malkan}, {Netzer}, \&
  {Peterson}}]{Reichert1994}
{Reichert}, G.~A., {Rodriguez-Pascual}, P.~M., {Alloin}, D., {et~al.} 1994,
  \apj, 425, 582

\bibitem[{{Shen}(2013)}]{Shen2013}
{Shen}, Y. 2013, Bulletin of the Astronomical Society of India, 41, 61

\bibitem[{{Shen} \& {Ho}(2014)}]{Shen2014}
{Shen}, Y., \& {Ho}, L.~C. 2014, \nat, 513, 210

\bibitem[{{Shen} \& {Kelly}(2010)}]{Shen2010}
{Shen}, Y., \& {Kelly}, B.~C. 2010, \apj, 713, 41

\bibitem[{{Shen} \& {Kelly}(2012)}]{Shen2012}
---. 2012, The Astrophysical Journal, 746, 169

\bibitem[{{Shen} {et~al.}(2011){Shen}, {Richards}, {Strauss}, {Hall},
  {Schneider}, {Snedden}, {Bizyaev}, {Brewington}, {Malanushenko},
  {Malanushenko}, {Oravetz}, {Pan}, \& {Simmons}}]{Shen2011}
{Shen}, Y., {Richards}, G.~T., {Strauss}, M.~A., {et~al.} 2011, \apjs, 194, 45

\bibitem[{{Shen} {et~al.}(2015{\natexlab{a}}){Shen}, {Brandt}, {Dawson},
  {Hall}, {McGreer}, {Anderson}, {Chen}, {Denney}, {Eftekharzadeh}, {Fan},
  {Gao}, {Green}, {Greene}, {Ho}, {Horne}, {Jiang}, {Kelly}, {Kinemuchi},
  {Kochanek}, {P{\^a}ris}, {Peters}, {Peterson}, {Petitjean}, {Ponder},
  {Richards}, {Schneider}, {Seth}, {Smith}, {Strauss}, {Tao}, {Trump},
  {Wood-Vasey}, {Zu}, {Eisenstein}, {Pan}, {Bizyaev}, {Malanushenko},
  {Malanushenko}, \& {Oravetz}}]{Shen2015a}
{Shen}, Y., {Brandt}, W.~N., {Dawson}, K.~S., {et~al.} 2015{\natexlab{a}}, The
  Astrophysical Journal Supplement Series, 216, 4

\bibitem[{{Shen} {et~al.}(2015{\natexlab{b}}){Shen}, {Greene}, {Ho}, {Brand t},
  {Denney}, {Horne}, {Jiang}, {Kochanek}, {McGreer}, {Merloni}, {Peterson},
  {Petitjean}, {Schneider}, {Schulze}, {Strauss}, {Tao}, {Trump}, {Pan}, \&
  {Bizyaev}}]{Shen2015b}
{Shen}, Y., {Greene}, J.~E., {Ho}, L.~C., {et~al.} 2015{\natexlab{b}}, The
  Astrophysical Journal, 805, 96

\bibitem[{{Shen} {et~al.}(2016{\natexlab{a}}){Shen}, {Horne}, {Grier},
  {Peterson}, {Denney}, {Trump}, {Sun}, {Brandt}, {Kochanek}, \&
  {Dawson}}]{Shen2016a}
{Shen}, Y., {Horne}, K., {Grier}, C.~J., {et~al.} 2016{\natexlab{a}}, \apj,
  818, 30

\bibitem[{{Shen} {et~al.}(2016{\natexlab{b}}){Shen}, {Brandt}, {Richards},
  {Denney}, {Greene}, {Grier}, {Ho}, {Peterson}, {Petitjean}, {Schneider},
  {Tao}, \& {Trump}}]{Shen2016b}
{Shen}, Y., {Brandt}, W.~N., {Richards}, G.~T., {et~al.} 2016{\natexlab{b}},
  \apj, 831, 7

\bibitem[{{Shen} {et~al.}(2019{\natexlab{a}}){Shen}, {Grier}, {Horne},
  {Brandt}, {Trump}, {Hall}, {Kinemuchi}, {Starkey}, {Schneider}, {Ho},
  {Homayouni}, {I-Hsiu Li}, {McGreer}, {Peterson}, {Bizyaev}, {Chen}, {Dawson},
  {Eftekharzadeh}, {Green}, {Guo}, {Jia}, {Jiang}, {Kneib}, {Li}, {Li}, {Nie},
  {Oravetz}, {Oravetz}, {Pan}, {Petitjean}, {Ponder}, {Rogerson}, {Vivek},
  {Zhang}, \& {Zou}}]{Shen2019b}
{Shen}, Y., {Grier}, C.~J., {Horne}, K., {et~al.} 2019{\natexlab{a}}, \apjl,
  883, L14

\bibitem[{{Shen} {et~al.}(2019{\natexlab{b}}){Shen}, {Hall}, {Horne}, {Zhu},
  {McGreer}, {Simm}, {Trump}, {Kinemuchi}, {Brandt}, {Green}, {Grier}, {Guo},
  {Ho}, {Homayouni}, {Jiang}, {I-Hsiu Li}, {Morganson}, {Petitjean},
  {Richards}, {Schneider}, {Starkey}, {Wang}, {Chambers}, {Kaiser},
  {Kudritzki}, {Magnier}, \& {Waters}}]{Shen2019a}
{Shen}, Y., {Hall}, P.~B., {Horne}, K., {et~al.} 2019{\natexlab{b}}, \apjs,
  241, 34

\bibitem[{{Smee} {et~al.}(2013){Smee}, {Gunn}, {Uomoto}, {Roe}, {Schlegel},
  {Rockosi}, {Carr}, {Leger}, {Dawson}, {Olmstead}, {Brinkmann}, {Owen},
  {Barkhouser}, {Honscheid}, {Harding}, {Long}, {Lupton}, {Loomis}, {Anderson},
  {Annis}, {Bernardi}, {Bhardwaj}, {Bizyaev}, {Bolton}, {Brewington}, {Briggs},
  {Burles}, {Burns}, {Castander}, {Connolly}, {Davenport}, {Ebelke}, {Epps},
  {Feldman}, {Friedman}, {Frieman}, {Heckman}, {Hull}, {Knapp}, {Lawrence},
  {Loveday}, {Mannery}, {Malanushenko}, {Malanushenko}, {Merrelli}, {Muna},
  {Newman}, {Nichol}, {Oravetz}, {Pan}, {Pope}, {Ricketts}, {Shelden},
  {Sandford}, {Siegmund}, {Simmons}, {Smith}, {Snedden}, {Schneider},
  {SubbaRao}, {Tremonti}, {Waddell}, \& {York}}]{Smee2013}
{Smee}, S.~A., {Gunn}, J.~E., {Uomoto}, A., {et~al.} 2013, \aj, 146, 32

\bibitem[{{Starkey} {et~al.}(2016){Starkey}, {Horne}, \&
  {Villforth}}]{Starkey2016}
{Starkey}, D.~A., {Horne}, K., \& {Villforth}, C. 2016, \mnras, 456, 1960

\bibitem[{{Sun} {et~al.}(2015){Sun}, {Trump}, {Shen}, {Brand t}, {Dawson},
  {Denney}, {Hall}, {Ho}, {Horne}, {Jiang}, {Richards}, {Schneider}, {Bizyaev},
  {Kinemuchi}, {Oravetz}, {Pan}, \& {Simmons}}]{Sun2015}
{Sun}, M., {Trump}, J.~R., {Shen}, Y., {et~al.} 2015, \apj, 811, 42

\bibitem[{{Trakhtenbrot} \& {Netzer}(2012)}]{Trakhtenbot2012}
{Trakhtenbrot}, B., \& {Netzer}, H. 2012, \mnras, 427, 3081

\bibitem[{{Trevese} {et~al.}(2007){Trevese}, {Paris}, {Stirpe}, {Vagnetti}, \&
  {Zitelli}}]{Trevese2007}
{Trevese}, D., {Paris}, D., {Stirpe}, G.~M., {Vagnetti}, F., \& {Zitelli}, V.
  2007, \aap, 470, 491

\bibitem[{{Vestergaard} \& {Osmer}(2009)}]{Vestergaard2009}
{Vestergaard}, M., \& {Osmer}, P.~S. 2009, \apj, 699, 800

\bibitem[{{Vestergaard} \& {Peterson}(2006)}]{Vestergaard2006}
{Vestergaard}, M., \& {Peterson}, B.~M. 2006, \apj, 641, 689

\bibitem[{{Wang} {et~al.}(2019){Wang}, {Shen}, {Jiang}, {Horne}, {Brandt},
  {Grier}, {Ho}, {Homayouni}, {I-Hsiu Li}, {Schneider}, \& {Trump}}]{Wang2019}
{Wang}, S., {Shen}, Y., {Jiang}, L., {et~al.} 2019, \apj, 882, 4

\bibitem[{{Williams} {et~al.}(2004){Williams}, {Olszewski}, {Lesser}, \&
  {Burge}}]{Williams2004}
{Williams}, G.~G., {Olszewski}, E., {Lesser}, M.~P., \& {Burge}, J.~H. 2004, in
  Society of Photo-Optical Instrumentation Engineers (SPIE) Conference Series,
  Vol. 5492, \procspie, ed. A.~F.~M. {Moorwood} \& M.~{Iye}, 787--798

\bibitem[{{Woo}(2008)}]{Woo2008}
{Woo}, J.-H. 2008, \aj, 135, 1849

\bibitem[{{Woo} {et~al.}(2015){Woo}, {Yoon}, {Park}, {Park}, \&
  {Kim}}]{Woo2015}
{Woo}, J.-H., {Yoon}, Y., {Park}, S., {Park}, D., \& {Kim}, S.~C. 2015, \apj,
  801, 38

\bibitem[{{Yu} {et~al.}(2020){Yu}, {Kochanek}, {Peterson}, {Zu}, {Brandt},
  {Cackett}, {Fausnaugh}, \& {McHardy}}]{Yu2020}
{Yu}, Z., {Kochanek}, C.~S., {Peterson}, B.~M., {et~al.} 2020, \mnras, 491,
  6045

\bibitem[{{Zu} {et~al.}(2011){Zu}, {Kochanek}, \& {Peterson}}]{Zu2011}
{Zu}, Y., {Kochanek}, C.~S., \& {Peterson}, B.~M. 2011, \apj, 735, 80

\end{thebibliography}

\end{document}